\documentclass{article}

\usepackage{natbib}
\setcitestyle{numbers,square}

\usepackage{amsmath} 
\usepackage{graphicx} 
\usepackage{makecell}

\usepackage[preprint]{neurips_2025}

\usepackage[utf8]{inputenc} 
\usepackage[T1]{fontenc}    
\usepackage{hyperref}       
\usepackage{url}            
\usepackage{booktabs}       
\usepackage{amsfonts}       
\usepackage{nicefrac}       
\usepackage{microtype}      
\usepackage{xcolor}         

\title{EPBench: A Benchmark for Short-term Earthquake Prediction with Neural Networks}

\author{%
Zhiyu Xu \\
\\
Jinan University\\
\\
\texttt{xuzhiyu@stu2022.jnu.edu.cn}
\And
  Qingliang Chen\thanks{corresponding author} \\
  \\
  Jinan University\\
   \\
  \texttt{tpchen@jnu.edu.cn} \\
}

\begin{document}

\maketitle

\begin{abstract}
   Since the beginning of this century, the significant  advancements in artificial intelligence and neural networks have offered the potential to bring new transformations to short-term earthquake prediction research. However, currently, there is  no widely used benchmark for this task. To address this, we have built a new benchmark (\textbf{EPBench}), which is, to our knowledge, the first global regional-scale short-term earthquake prediction benchmark. Our benchmark comprises 924,472 earthquake records and 2959 multimodal earthquake records collected from seismic networks around the world. Each record includes basic information such as time, longitude and latitude, magnitude, while each multimodal record includes waveform and moment tensor information additionally, covering a time span from 1970 to 2021. To evaluate  performance of models on this task, we have established a series of data partitions and evaluation methods tailored to the short-term earthquake prediction task. We also provide a variety of tools to assist future researchers in partitioning the data according to their  geographical understanding. Our benchmark includes a variety of neural network models widely used for time series forecasting, as well as a statistical-based model currently employed by seismological bureaus in several countries. We hope this benchmark will serve as a guide to attract more researchers to explore new methods for addressing this task, which holds great significance for human existence. Code is available at
   https://github.com/CoderZY-X/EPBench
\end{abstract}

\section{Introduction}
Earthquakes, along with the tsunamis and volcanic activities they trigger, claim tens of thousands of lives each year and result in massive economic losses. Developing viable and reliable earthquake prediction methods has long been a goal of humanity. The study of earthquakes dates back at least two millennia, with early research documented in ancient Greece and ancient China~\cite{pararas2011earthquake,feng2010research}.

In seismology, earthquake prediction can be divided into three categories: long-term, medium-term, and short-term prediction. Short-term earthquake prediction typically spans a few months or even weeks, while medium-term and long-term predictions cover much longer timeframes, usually predicting possible earthquakes that may occur over several years. As human's understanding of plate tectonics and plate movements has deepened, humanity has gained a better understanding of the causes of earthquakes and the regions prone to seismic activity, known as seismic belts.

Despite significant progress in long-term and medium-term earthquake prediction, humanity still lacks effective methods for short-term earthquake prediction, which is critically important. The only successful short-term earthquake prediction to date occurred in 1975 in Haicheng, China~\cite{wang2006predicting}. This earthquake struck a densely populated area, but fortunately, due to the successful prediction and subsequent evacuation, casualties were minimized, saving tens of thousands of lives. This achievement quickly caused a huge sensation in the global earthquake research community. However, despite numerous attempts, this method has not been successfully replicated, and to this day, there is still no effective and reliable method for humans to predict earthquakes that may occur within a month.

Since the beginning of the 21st century,  significant advancements in artificial intelligence and neural networks have revolutionized various scientific domains~\cite{bi2022pangu,jumper2021highly}. Concurrently, researchers have increasingly explored the application of these techniques to transform earthquake research ~\cite{chen2020neural,stockman2024earthquakenpp}. However, to our knowledge, there is still no widely recognized dataset or benchmark in the short-term prediction field, substantially impeding progress in this task.

To bridge this gap, we present the first global regional-scale short-term
earthquake prediction benchmark containing 924,472 earthquake records and 2959 multimodal earthquake records collected from seismic networks around the world.  Each record contains fundamental attributes including timestamp, geographic coordinates (longitude and latitude), and magnitude, while each multimodal record additionally incorporates waveform and moment tensor information, covering a global geographic span and a time span from 1970 to 2021. The multimodal records have been systematically organized by the USGS since March 1996, and we have obtained nearly all available multimodal data containing waveform and moment tensor information. For comprehensive evaluation, We select various foundational models commonly used in time series analysis, including CNN, RNN, Transformer-based, and diffusion-based models. While emphasizing neural network approaches, we also include a statistical-based model which is currently employed by seismological bureaus in several countries, known as the epidemic-type aftershock sequence model(ETAS). We also draw inspiration from the MSE metric commonly applied to various machine learning tasks, incorporating spatio-temporal characteristics and consider the false alarms (False Negative) and the omissive reports (False Positive), which are crucial for evaluating earthquake prediction models. These led us to establish a set of intuitive and effective metrics tailored for the earthquake prediction task for assessing performance of models and build our benchmark.

Additionally, we conducted preliminary regional and temporal partitioning based on our understanding and preprocessed the moment tensor information, detailed in Section 3 and Section 5. Experiments validated the effectiveness of these processes, revealing the potential for combining physical understanding and various physical characteristics to enhance model performance. Notably, we also provide tools to help researchers partition regions based on their own physical and geographical understanding. Please refer to our code.

In summary, our contributions can be summarized as follows:
\begin{itemize}
    \item We collected 924,472 earthquake  records and 2959 multimodal earthquake records from worldwide seismic networks, providing the potential to predict earthquakes using various physical characteristics. Additionally, we analyzed the dataset (please refer to the appendixes) and offered explanations of  different physical quantities to help researchers quickly understand the task and dataset.
    \item We select various foundational models commonly used in time series analysis along with a statistical-based model, and propose a set of intuitive and effective metrics tailored for the earthquake prediction task to comprehensively construct our benchmark. 
    \item We conducted preliminary regional and temporal partitioning based on our understanding and proposed methods to preprocessed the moment tensor information, with experiments validated the effectiveness of these processes. We also provided tools to help researchers partition regions  based on their own physical and geographical understanding.
\end{itemize}

\section{Related Works}
\textbf{Earthquake Prediction Benchmark} Chen et al.~\cite{chen2020neural} introduced an earthquake dataset to evaluate their proposed spatio-temporal point process model using neural ODEs, which includes earthquake records in Japan from 1990 to 2020 with magnitudes above 2.5, splitting the data into month-long segments with a 7-day offset. Stockman et al.~\cite{stockman2024earthquakenpp} proposed an earthquake prediction benchmark (EarthQuakeNPP) that includes earthquakes in California from 1971 to 2021, with models that solely consist of neural point process models. However, both their datasets and benchmarks have the following limitations: They have a small spatial span, and the amount of data is limited; They do not focus on short-term prediction; They do not use elaborate metrics for the earthquake prediction task instead simply replicate metrics for point process modeling; They focus exclusively on neural point process models, significantly overlooking other types of neural networks. 

\textbf{Neural Networks for Earthquake Research} Mousavi et al.~\cite{mousavi2019cred} introduced a CNN-RNN earthquake detector for seismic signal identification. Wang et al.~\cite{wang2017earthquake} designed a two-dimensional LSTM network to learn the spatio-temporal relationships among earthquakes at different locations. Wu et al.~\cite{wu2018deepdetect} proposed a novel cascaded region-based convolutional neural network to capture earthquake events, and Ross et al.~\cite{ross2019phaselink} developed PhaseLink, a framework based on recent advances in deep learning for grid-free earthquake phase association. These studies highlight the growing adoption of neural networks in seismology, underscoring the potential of artificial intelligence to advance earthquake research.

\textbf{Time Series Analysis Models} \ \ \textbf{ Transformer-based models} have revolutionized various fields beyond their initial NLP applications, primarily due to their attention mechanism, which effectively captures global, long-range dependencies in data.  These models are well-suited for handling both spatial and temporal dependencies, as demonstrated in applications like ST-LLM~\cite{liu2024spatial} for traffic prediction and STEP~\cite{shao2022pre}, which combines spatial information with temporal modeling capabilities. \textbf{Multi-Layer Perceptrons (MLPs)} are valued for their lightweight design, making them computationally efficient and cost-effective. Examples like TSMixer~\cite{ekambaram2023tsmixer} and TTMs~\cite{ekambaram2401ttms} demonstrate superior memory usage and processing speed while maintaining competitive performance. \textbf{Convolutional Neural Networks (CNNs)} excel in modeling spatial and temporal data, particularly in self-supervised learning for general time series. Architectures such as ResNet~\cite{dong2023simmtm} and dilated convolution layers have been foundational in this domain, utilizing predominantly 1D convolutional operations. TimesNet~\cite{wu2022timesnet} innovatively converts 1D time series data into 2D tensors, enhancing the identification of multi-periodicity and complex temporal variations through a parameter-efficient inception block. \textbf{Recurrent Neural Networks (RNNs)} are recognized for their effectiveness in handling sequential temporal data, with recent developments challenging Transformer-based models. The RWKV-TS~\cite{hou2024rwkv} model exemplifies this trend, offering resource-efficient modeling capable of managing longer sequences. Lastly, \textbf{Diffusion models}, while prominently applied in computer vision, have recently begun to be explored in time series analysis. These models learn complex data distributions by gradually adding and reversing noise, which allows them to tackle prediction, imputation, and anomaly detection effectively, capturing both spatial and temporal dynamics in applications like traffic forecasting~\cite{peebles2023scalable,wen2023diffstg}. Our benchmark includes these various models, making it comprehensive and multi-strategy, thus providing a foundation for further research on these fundamental models.

\section{Background and Problem Setup}
\subsection{Causes of Earthquake}
\begin{figure}[t]
    \centering
    \includegraphics[width=0.8\linewidth,height=5.5cm]{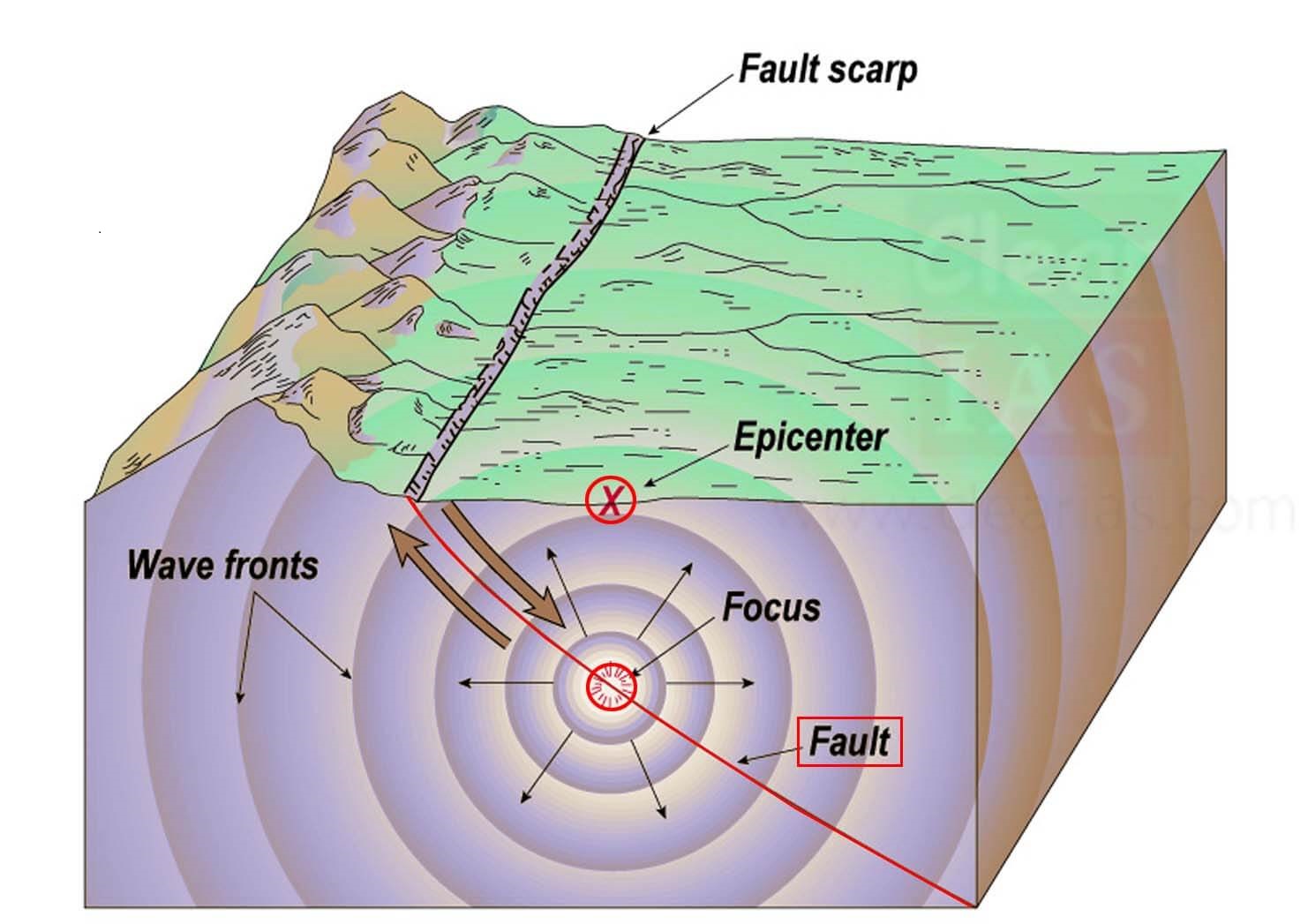}
    \caption{An illustration of earthquake occurrence}
    \label{fig:Figure 1}
\end{figure}

Earthquakes are primarily caused by the movement of tectonic plates within the Earth's crust. As these plates interact—colliding, sliding past each other, or pulling apart—stress accumulates at fault lines where the plates meet. This stress gradually increases until it exceeds the strength of the rocks, causing a sudden release of energy. When the crust breaks, it generates seismic waves that propagate through the Earth, leading to ground vibrations. See Fig.~\ref{fig:Figure 1}.

The focus (hypocenter) is the subsurface origin point where an earthquake originates. The epicenter is the point on the Earth's surface that is directly above the focus. Focal depth refers to the distance from the epicenter to the focus, which plays a crucial role in determining the intensity and effects of the earthquake. Generally, earthquakes with shallow focal depths (less than 70 km) tend to cause more damage at the surface, while those that occur at greater depths (over 300 km) usually have less impact. More information about the causes of earthquakes can be found in the appendixes.

\subsection{Problem Setup}
Given m historical earthquake event record vectors within a period of time, $ \mathbf{e}^1,\mathbf{e}^2,\cdot \cdot \cdot ,\mathbf{e}^m \ \in \mathbb{R}^N $, where $N$ represents the number of types of information regarding the event records, the goal is to predict the set of earthquake vectors $\hat{E}  =  \{ \mathbf{e}^{m+1},\mathbf{e}^{m+2},\cdot \cdot \cdot ,\mathbf{e}^{m+k}\}$ for the next time period. The primary predicted set of earthquake vectors can then serve as input for subsequent predictions of future earthquakes.

\begin{figure}[t]
    \centering
    \includegraphics[width=1.0\linewidth,height=7cm]{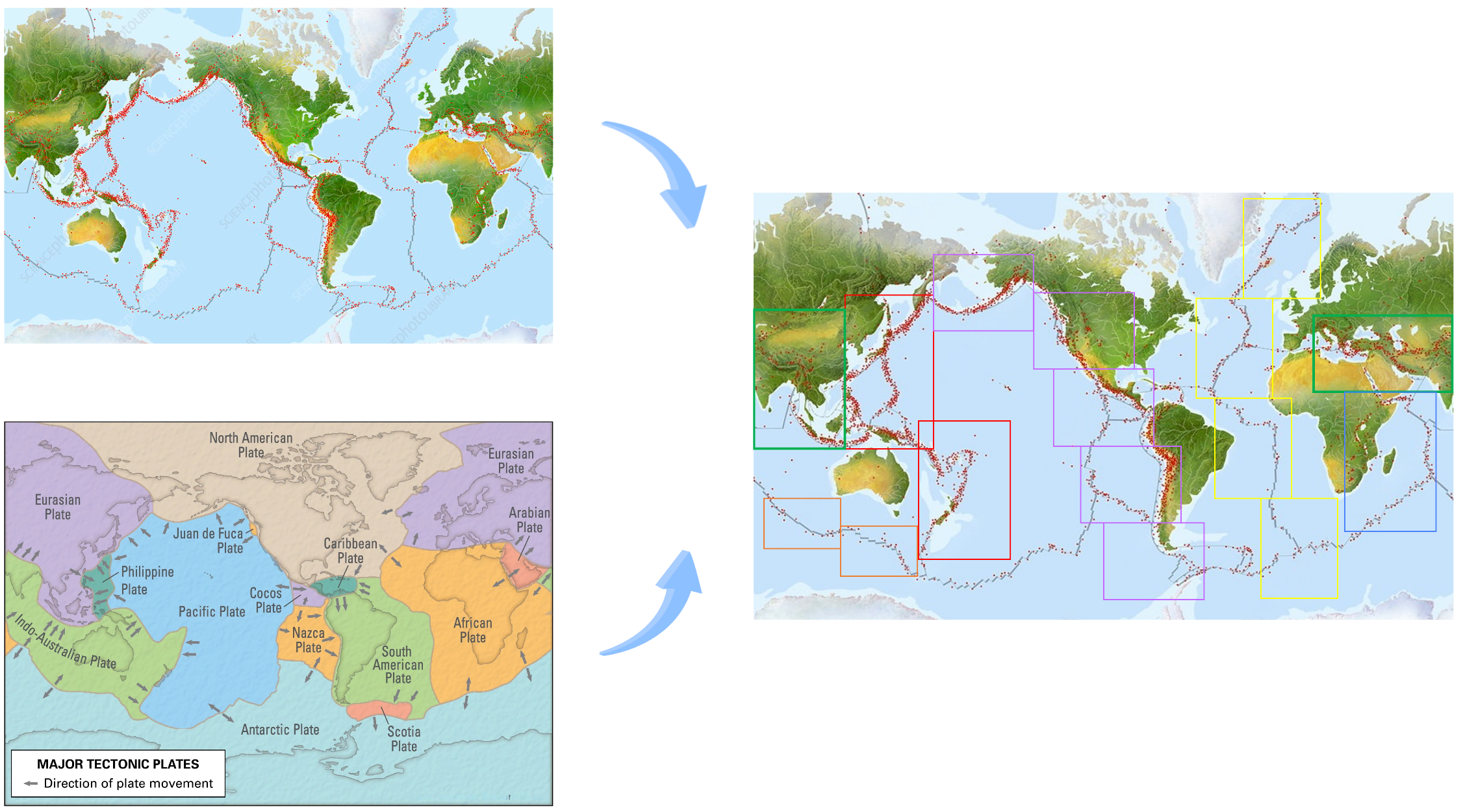}
    \caption{An illustration of our regional division. The boxes of the same color represent the same region. We assign simple names to each region based on the abbreviations of the countries or geographic landmarks within them. The green region is named "EU-CHN", the red region is named "JPN-IDN", the orange region is named "AU South", the purple region is named "USA-CL", the yellow region is named "Atlantic", and the blue region is named "AfricaAsia".}
    \label{fig:Figure 2}
\end{figure}

We have divided the earthquake records into regions based on the spatial distribution of earthquake occurrences and the tectonic plates of the Earth with the trends of their movements, shown in Fig.~\ref{fig:Figure 2}. Our main idea is to group together earthquakes that occur on the same boundary or are caused by similar movement trends, as this may help to better analyze the occurrence trends of earthquakes. Each region is trained, predicted, and evaluated separately, and the experimental results have demonstrated the effectiveness of our division. Moreover, we also provide tools in our code to assist future researchers in dividing regions based on better physical and seismological understanding, making the regional division more rational.

\section{EPBench}
\subsection{Dataset}
We obtained data from the NSF Seismological Facility for the Advancement of Geoscience (SAGE) and  the U.S. Geological Survey (USGS), which compile information from seismic networks of various countries. Our dataset covers global seismic events with magnitudes above 2.5 from 1970 to 2020, including key information such as latitude, longitude, time, magnitude, and depth. Furthermore, with the exploration of earthquake focal mechanisms by geoscientists in the late $20^{th}$ century, the moment tensor information of earthquakes with magnitudes of six and above has been analyzed and recorded regularly. Additionally, the enormous energy released by earthquakes of this magnitude results in reliable waveform data that is robust against noise, so we also included these in our dataset.

In summary, our dataset consists of two categories of data: one category includes all recorded earthquakes with magnitudes above 2.5 from 1970 to 2020 containing key information such as latitude, longitude, time, magnitude and depth; the other category includes  multimodal data with magnitudes above 6 from 1996 to 2021, which not only contains key information but also physical quantities such as waveform and  moment tensors. Seismologists aim to better understand earthquakes by analyzing the focal mechanisms of high-magnitude earthquakes, and we also hope to improve earthquake prediction by leveraging this multimodal information and injecting physical knowledge into the models. Both categories of data are global in scope. Below is a brief explanation of the various types of information contained within the seismic data:
\begin{itemize}
  \item  Longitude and Latitude: These two quantities specify the location of an earthquake, similar to $\theta$  and  $\varphi$ coordinates in a spherical coordinate system. They are recorded with a precision of four decimal places in our dataset.
    \item  Time: Indicates the occurrence time of the earthquake, accurate to milliseconds in our data.
    \item Magnitude: An indicator of the energy released by an earthquake, the larger the magnitude, the more intense the earthquake will be. Our magnitude data are measured on the Richter scale.
    \item  Depth(Focal Depth): The distance from the epicenter to the  focus.
    \item Moment Tensor: A mathematical representation that describes the distribution of stress and forces involved in earthquakes, providing insights into the earthquakes' focal mechanisms and the way they deform the surrounding materials.
    \item Waveform: The graphical representation of seismic waves generated by an earthquake, which includes information such as amplitude, frequency, and timing.
\end{itemize}
Fig.3 is an illustration of these  multimodal information. More detailed explanation of these physical quantities can be found in the appendixes.

\subsection{Dataset Splitting}
We divided our first category of data which covers a 50-year span, into two groups for training and evaluation. The first group uses all earthquake records from January 1, 1970, to January 1, 1995, for training, and then predicts earthquakes occurring in the next three months (from January 1995 to March 1995), evaluated using real earthquake records. The second group uses data from January 1, 1995, to January 1, 2020, for training, and predicts earthquakes in the following three months (from January 2020 to March 2020).

For our second category of multimodal data which covers a 25-year span, we use the entire dataset from March 1, 1996, to March 1, 2021 as the training data for a control experiment, then predict earthquakes occurring in a year (from March 2021 to March 2022). The evaluation is conducted using real earthquakes records  of magnitude 6 or above. The first experiment only uses basic information (time, longitude, latitude, magnitude) for training, while the second experiment incorporates moment tensor information. We take the cosine encoding of all angular information from the moment tensors and sequentially place each moment tensor angular information into separate channels with batch normalization applied. The experimental results in Table ~\ref{Table 3} demonstrate the effectiveness of our preprocessing.

In addition to the aforementioned two experiments, to validate the effectiveness of our global seismic region division, we selected three regions and further partitioned them into sub-regions, performing testing exclusively on data in these sub-regions while using training data from the regions specified in the first line of each experiment in Table ~\ref{Table 1}. As demonstrated, our approach effectively aggregates seismic events originating based on the tectonic plate and trend, thereby providing models with more comprehensive information.
\subsection{Metrics}
\begin{figure}[t]
    \centering
    \includegraphics[width=0.88\linewidth,height=7cm]{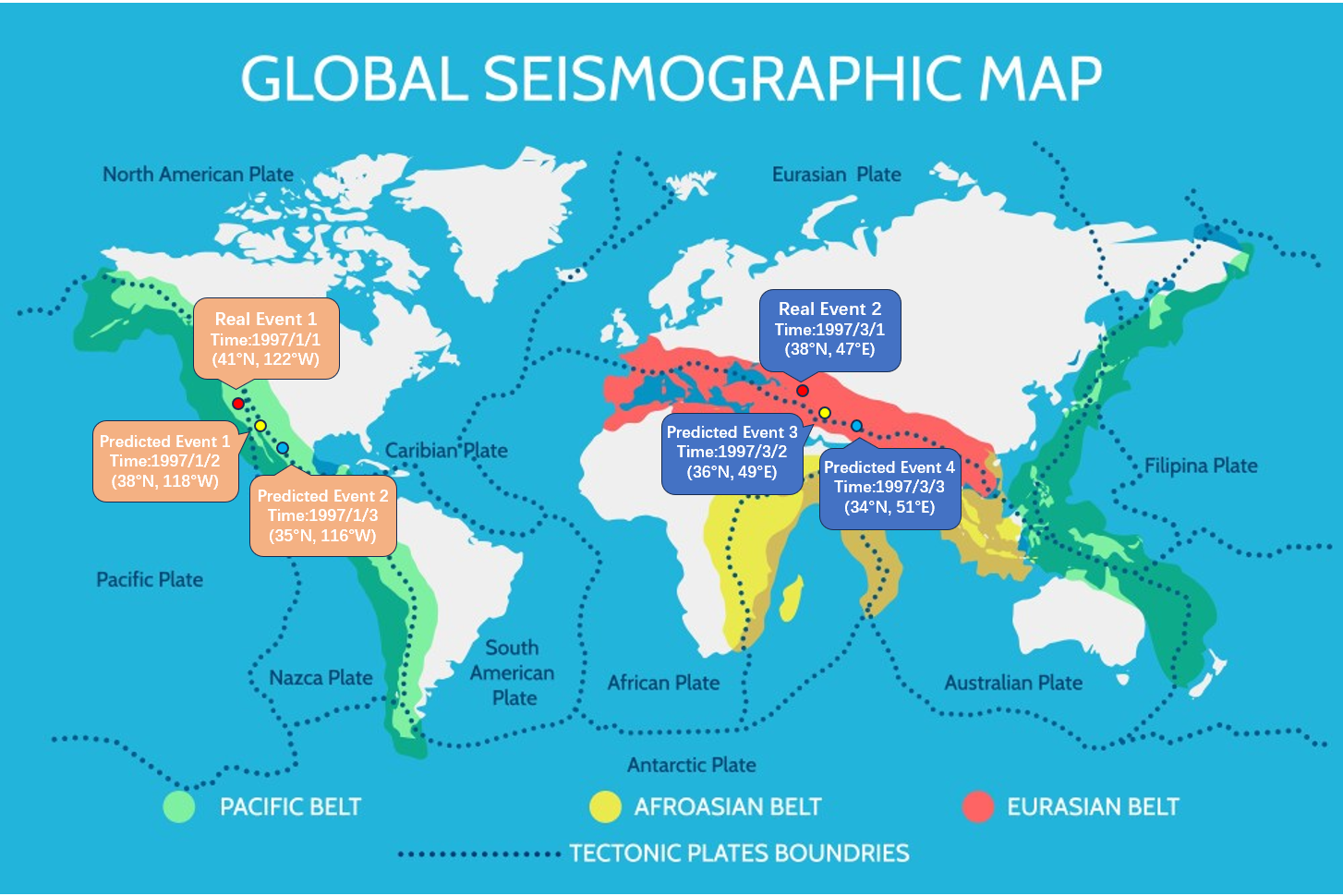}
    \caption{Examples of predicted situations. Note that, although neither the yellow nor the blue model accurately predicted the real earthquakes, the performance of the yellow model should be considered to be better than that of the blue model.}
    \label{fig:Figure 3}
\end{figure}

For earthquakes, especially high-magnitude earthquakes, which are discrete events that significantly impact humans, both the accuracy of prediction and the false alarm rate are very important. Improving the accuracy of prediction can help humanity anticipate earthquakes and respond, while reducing the false alarm rate can prevent a significant waste of resources. Both aspects have substantial benefits. At the same time, many existing common metrics for evaluation are not well-suited for earthquake forecasting: we cannot simply categorize predictions that do not completely match as incorrect, as shown in Figure ~\ref{fig:Figure 3}.

Therefore, we referenced True Positive(TP) and False Negative(FN) and proposed metrics (\textbf{Matching Rate} and \textbf{False Alarm Rate}) for measuring prediction accuracy and false alarm rate. Additionally, we referenced the Mean Squared Error (MSE) to develop metrics for assessing temporal and spatial discrepancies. Our core idea is to introduce spatio-temporal windows and aggregate real events and predicted events by calculating their spatio-temporal discrepancies, thus enabling a more reasonable evaluation of model performance.

First, for measuring the temporal discrepancies, we take the absolute value of the discrepancies between the occurrence times of two earthquake events. 

For measuring the spatial discrepancy, we use the Haversine formula to calculate the distance between the locations of  two earthquake events. The Haversine formula is widely used in geographic measurements and navigation, and it computes the distance between two points on the surface of a sphere based on their latitude and longitude, as follows:
\begin{equation}
    d = 2R \arcsin\left( \sqrt{\sin^2\left(\frac{\Delta lat}{2}\right) + \cos(lat_1) \cdot \cos(lat_2) \cdot \sin^2\left(\frac{\Delta lon}{2}\right)} \right) 
\end{equation}
Where $\Delta lat=lat_1-lat_2$ represents the latitude discrepancy, $\Delta lon=lon_1-lon_2$ represents longitude discrepancy and $R$ represents the radius length of the Earth, usually around 6371 kilometers. The proof of Haversine formula can be found in the appendixes. 

We define a spatio-temporal window for each real earthquake event with a magnitude above 4.5, which includes a time window ($T_w$) of 2 days and a distance window ($D_W$) of 150 km. If any predicted event falls within a time discrepancy of no more than 2 days and a spatial discrepancy of no more than 150 km from the real event, we consider it a proper match. The ratio of properly matched real events to the total number of real events is named the \textbf{Matching Rate(MR)}, which is analogous to True Positive(TP). Conversely, for each predicted event with a magnitude above 4.5, if no real earthquake with a magnitude above 4.5 occurs within this spatio-temporal window, we consider it a false alarm. The ratio of false alarm predictions to the total number of predictions is called the \textbf{False Alarm Rate(FAR)}, which is analogous to False Negative(FN).
\begin{equation}
    \text{MR} = \frac{
    \sum\limits_{i=1}^{N_{\text{real}}} 
    \mathbb{I}\left( 
        \exists \, j \, \text{ s.t. } 
        |t_{\text{real},i} - t_{\text{pred},j}| \leq T_w \,  
        \, \cap \, 
        d(\mathbf{x}_{\text{real},i}, \mathbf{x}_{\text{pred},j}) \leq D_w \,  
    \right)
}{
    N_{\text{real}}
}
\end{equation}
\begin{equation}
    \text{FAR} = \frac{
    \sum\limits_{j=1}^{N_{\text{pred}}} 
    \mathbb{I}\left( 
        \forall \, i, \, 
        |t_{\text{pred},j} - t_{\text{real},i}| > T_w \,  
        \, \cup \, 
        d(\mathbf{x}_{\text{pred},j}, \mathbf{x}_{\text{real},i}) > D_w \, 
    \right)
}{
    N_{\text{pred}}
}
\end{equation}
Where $\mathbb{I}()$ is the indicator function that equals 1 if the condition is satisfied and 0 otherwise. Time($t$) is measured in days, $d()$ represents the Haversine formula for calculating the distance between two locations, measured in kilometers. $N_{real}$ 
denotes the total number of real earthquake events with magnitudes above 4.5 and $N_{pred}$  denotes the total number of predicted earthquake events with magnitudes above 4.5.

After calculating the Matching Rate (MR) and False Alarm Rate (FAR), we further evaluate the spatio-temporal discrepancy between predicted and real events by introducing the Spatio-temporal Mean Squared Error (ST-MSE). We use the size of the spatio-temporal window we set as a criterion for measurement, weighting temporal and spatial discrepancies. Additionally, we normalize the range of the ST-MSE to [0, 1]. Specifically, The calculation formula for ST-MSE is as follows:
\begin{equation}
    \text{ST-MSE} = \frac{1}{2} \left( \left( \frac{t_{\text{pred}} - t_{\text{real}}}{T_w} \right)^2 + \left( \frac{d(\mathbf{x}_{\text{pred}}, \mathbf{x}_{\text{real}})}{D_w} \right)^2 \right)
\
\end{equation}
Since we only calculate the ST-MSE of predicted events within the spatio-temporal window, the range of ST-MSE is [0, 1], the same as MSE. We select, for each real event, the predicted event with the minimum ST-MSE as its only match. If no predicted events fall within this spatio-temporal window, the real event is considered a prediction failure. After we calculate the MR and FAR, we use the average minimum ST-MSE of all matched real events as a metric to quantify the spatio-temporal discrepancy between predicted and real events. It’s important to note that, ST-MSE excludes real events with prediction failure, as we have already used MR as a prior metric to evaluate the model's prediction success rate.

As mentioned in Section 4.1, the magnitude is an indicator of the energy released by an earthquake. For each real event, we also include the magnitude discrepancy of its matched predicted event.
\begin{equation}
    \Delta mag = | \ mag_{real}-mag_{pred} \ |
\
\end{equation}

\section{Baselines on EPBench}
We constructed the main baselines by combining a famous statistical seismology method and various time series foundational neural network models. Specifically, we adopted: (1) the epidemic-type aftershock sequence model (ETAS), which is based on statistics and widely used by various national seismological bureaus; (2) various neural network models for time series analysis including CNN, RNN, Transformer-based, and diffusion-based models.

The ETAS model we use is implemented in
the etas python package by Mizrahi et al.~\cite{mizrahi2021embracing}. Its form is as follows:
\begin{equation}  
\lambda(t, \mathbf{x} \mid \mathcal{H}_t ; \theta) = \mu + \sum_{i : t_i < t} g(t - t_i, \ \| \mathbf{x} - \mathbf{x}_i \|_2^2, m_i),  
\end{equation}  
\begin{equation}  
g(t, r^2, m) = \frac{e^{-t/\tau} \cdot k \cdot e^{a(m-M_c)}}{(t+c)^{1+\omega} \cdot (r^2 + d \cdot e^{\gamma(m-M_c)})^{1+\rho}} 
\end{equation} 
Where $\mathcal{H}_t$ denotes the history of the events preceding time $t$, $\theta$  is a set of parameters that can be iteratively optimized. $r^2$ is the squared distance between events, and $k,a,c,\omega,\tau,d,\gamma,\rho$ are learnable parameters along with the constant background rate $\mu$.

Based on previous research where various neural network models were developed for earthquake prediction and time series analysis~\cite{wang2017earthquake,wen2023diffstg,dong2023simmtm}, We designed four different neural network models, including EP-CNN, EP-RNN, EP-Transformer, and EP-Diff. The specific structure of them and the implementation details can be found in the appendixes.

The experimental results are shown in Table ~\ref{Table 2}. As can be seen from the table, although the ETAS model achieves an MR of over 90\% in the vast majority of cases, its FAR  is also as high as 90\%. In other words, the ETAS employs a strategy of predicting an extremely large number of earthquakes to cover potential seismic events. However, the excessively high 90\% FAR severely limits the practical application of the ETAS model, as the enormous number of false alarms makes it unfeasible in terms of reliability. In contrast, although the neural network models do not achieve high  MR, their FAR are significantly lower than that of ETAS. Based on the current results, if we aim to develop an effective and reliable earthquake prediction model, we need to either substantially reduce the FAR of ETAS or improve the MR of the neural network models—while maintaining FAR at a low level.

\begin{table}[h]
\centering
\caption{Quantitative results of region division. Note that the second column group represents subregions of the regions indicated in the first column group, and in the "Global" column group, we use global data for training. The training data for the three groups are different, but all tests are conducted only on the smallest regions indicated in the second column group.}
 \scalebox{0.66}[0.81]{
\begin{tabular}{|l|llll|llll|llll|}
\hline
               & \multicolumn{4}{l|}{\makecell[c]{EU-CHN(1995$ \sim $2020)} }              & \multicolumn{4}{l|}{\makecell[c]{CHN(1995$ \sim $2020)}}                  & \multicolumn{4}{l|}{\makecell[c]{Global(1995$ \sim $2020)}}               \\ \cline{2-13} 
               & \makecell[c]{MR}     & \makecell[c]{ST-MSE} & \multicolumn{1}{l|}{\makecell[c]{$\Delta$mag}}  & \makecell[c]{FAR}    & \makecell[c]{MR}     & \makecell[c]{ST-MSE} & \multicolumn{1}{l|}{\makecell[c]{$\Delta$mag}}  & \makecell[c]{FAR}    & \makecell[c]{MR }    & \makecell[c]{ST-MSE} & \multicolumn{1}{l|}{\makecell[c]{$\Delta$mag}}  & \makecell[c]{FAR}    \\ \hline
EP-LSTM        & \makecell[c]{5.68\%}  & \makecell[c]{0.23}   & \multicolumn{1}{l|}{\makecell[c]{0.49}}                      & \makecell[c]{91.85}\% & \makecell[c]{3.62}\%  & \makecell[c]{0.31}   & \multicolumn{1}{l|}{\makecell[c]{0.41}}                      & \makecell[c]{93.15}\% & \makecell[c]{0\%}     & \makecell[c]{-}      & \multicolumn{1}{l|}{\makecell[c]{-}}                         & \makecell[c]{100\%}   \\
EP-Transformer & \makecell[c]{9.52}\%  & \makecell[c]{0.45}   & \multicolumn{1}{l|}{\makecell[c]{0.37}}                      & \makecell[c]{96.52\%} & \makecell[c]{5.66\%}  & \makecell[c]{0.19}   & \multicolumn{1}{l|}{\makecell[c]{0.22}}                      & \makecell[c]{94.18\%} & \makecell[c]{3.66\%}  & \makecell[c]{0.35}   & \multicolumn{1}{l|}{\makecell[c]{0.81}}                      & \makecell[c]{98.12\%} \\
EP-Diff        & \makecell[c]{12.65\%} & \makecell[c]{0.42}   & \multicolumn{1}{l|}{\makecell[c]{0.73}}                      & \makecell[c]{89.44\%} & \makecell[c]{9.02\%}  & \makecell[c]{0.27}   & \multicolumn{1}{l|}{\makecell[c]{0.63}}                      & \makecell[c]{90.03\%} & \makecell[c]{6.92\%}  & \makecell[c]{0.58}   & \multicolumn{1}{l|}{\makecell[c]{0.45}}                      & 96.55\% \\ \hline
               & \multicolumn{4}{l|}{\makecell[c]{JPN-IDN(1995$ \sim $2020)}}              & \multicolumn{4}{l|}{\makecell[c]{IDN(1995$ \sim $2020)}}                  & \multicolumn{4}{l|}{\makecell[c]{Global(1995$ \sim $2020)}}               \\ \cline{2-13} 
               & \makecell[c]{MR}     & \makecell[c]{ST-MSE} & \multicolumn{1}{l|}{\makecell[c]{$\Delta$mag}}  & \makecell[c]{FAR}    & \makecell[c]{MR }    & \makecell[c]{ST-MSE} & \multicolumn{1}{l|}{\makecell[c]{$\Delta$mag}}  & \makecell[c]{FAR}    & \makecell[c]{MR}     & \makecell[c]{ST-MSE} & \multicolumn{1}{l|}{\makecell[c]{$\Delta$mag}}                       & \makecell[c]{FAR}    \\ \hline
EP-LSTM        & \makecell[c]{7.42\%}  & \makecell[c]{0.31}   & \multicolumn{1}{l|}{\makecell[c]{0.46}}                      & \makecell[c]{93.43\%} & \makecell[c]{4.98\%}  & \makecell[c]{0.42}   & \multicolumn{1}{l|}{\makecell[c]{0.30}}                      & \makecell[c]{92.28\%} & \makecell[c]{4.34\%}  & \makecell[c]{0.49}   & \multicolumn{1}{l|}{\makecell[c]{0.61}}                      & \makecell[c]{96.37\%} \\
EP-Transformer & \makecell[c]{12.43\%} &\makecell[c]{ 0.52}   & \multicolumn{1}{l|}{\makecell[c]{0.38}}                      & \makecell[c]{90.28\%} & \makecell[c]{9.77\%}  & \makecell[c]{0.56}   & \multicolumn{1}{l|}{\makecell[c]{0.57}}                      & \makecell[c]{93.44\%} & \makecell[c]{7.55\%}  & \makecell[c]{0.67}   & \multicolumn{1}{l|}{\makecell[c]{0.48}}                      & \makecell[c]{95.82\%} \\
EP-Diff        & \makecell[c]{16.91\%} & \makecell[c]{0.44}   & \multicolumn{1}{l|}{\makecell[c]{0.47}}                      & \makecell[c]{87.45\%} & \makecell[c]{12.88\%} & \makecell[c]{0.71}   & \multicolumn{1}{l|}{\makecell[c]{0.41}}                      & \makecell[c]{91.31\%} & \makecell[c]{9.61\%}  & \makecell[c]{0.54}   & \multicolumn{1}{l|}{\makecell[c]{0.49}}                      & \makecell[c]{92.26\%} \\ \hline
               & \multicolumn{4}{l|}{\makecell[c]{USA-CL(1995$ \sim $2020)}}                          & \multicolumn{4}{l|}{\makecell[c]{USA(1995 $ \sim $2020)}}                  & \multicolumn{4}{l|}{\makecell[c]{Global(1995$ \sim $2020)}}               \\ \cline{2-13} 
               & \makecell[c]{MR}     & \makecell[c]{ST-MSE} & \multicolumn{1}{l|}{\makecell[c]{$\Delta$mag}}  & \makecell[c]{FAR}    & \makecell[c]{MR }    & \makecell[c]{ST-MSE} & \multicolumn{1}{l|}{\makecell[c]{$\Delta$mag}}  & \makecell[c]{FAR}    & \makecell[c]{MR }    & \makecell[c]{ST-MSE} & \multicolumn{1}{l|}{\makecell[c]{$\Delta$mag}}  & \makecell[c]{FAR}    \\ \hline
EP-LSTM        & \makecell[c]{8.90\%}  & \makecell[c]{0.25}   & \multicolumn{1}{l|}{\makecell[c]{0.91}} & \makecell[c]{92.21\%} & \makecell[c]{7.69\%}  & \makecell[c]{0.35}   & \multicolumn{1}{l|}{\makecell[c]{0.47}} & \makecell[c]{97.25\%} & \makecell[c]{8.01\%}  & \makecell[c]{0.68}   & \multicolumn{1}{l|}{\makecell[c]{0.74}} & \makecell[c]{92.64\%} \\
EP-Transformer & \makecell[c]{18.62\%} & \makecell[c]{0.54}   & \multicolumn{1}{l|}{\makecell[c]{0.64}} & \makecell[c]{82.66\%} & \makecell[c]{13.73\%} & \makecell[c]{0.51}   & \multicolumn{1}{l|}{\makecell[c]{0.76}} &\makecell[c]{ 91.32\%} & \makecell[c]{15.79\%} & \makecell[c]{0.56}   & \multicolumn{1}{l|}{\makecell[c]{0.25}} & \makecell[c]{88.57\% }\\
EP-Diff        & \makecell[c]{22.58\%} & \makecell[c]{0.38}   & \multicolumn{1}{l|}{\makecell[c]{0.83}} & \makecell[c]{75.76\%} & \makecell[c]{16.53\%} & \makecell[c]{0.23}   & \multicolumn{1}{l|}{\makecell[c]{1.04}} & \makecell[c]{88.30\%} & \makecell[c]{17.11\%} & \makecell[c]{0.49}   & \multicolumn{1}{l|}{\makecell[c]{0.84}} & \makecell[c]{80.09\%} \\ \hline
\end{tabular}}
\label{Table 1}
\end{table}

\begin{table}[t]
\centering
\caption{Evaluating performance of our baseline models on EPBench data.}
 \scalebox{0.6}[0.6]{
\begin{tabular}{|l|cccc|cccc|}
\hline
               & \multicolumn{4}{c|}{EU-CHN(1975$\sim$1995)}                 & \multicolumn{4}{c|}{EU-CHN(1995$\sim$2020)}                 \\ \cline{2-9} 
               & MR      & ST-MSE & \multicolumn{1}{c|}{$\Delta$mag}  & FAR     & MR      & ST-MSE & \multicolumn{1}{c|}{$\Delta$mag}  & FAR     \\ \hline
ETAS           & 95.12\% & 0.10   & \multicolumn{1}{c|}{1.57} & 97.12\% & 93.31\% & 0.11   & \multicolumn{1}{c|}{1.28} & 98.45\% \\
EP-CNN         & 1.63\%  & 0.25   & \multicolumn{1}{c|}{0.25} & 99.70\% & 4.91\%  & 0.71   & \multicolumn{1}{c|}{0.57} & 98.96\% \\
EP-LSTM        & 6.05\%  & 0.36   & \multicolumn{1}{c|}{0.81} & 98.54\% & 8.48\%  & 0.54   & \multicolumn{1}{c|}{0.59} & 97.31\% \\
EP-Transformer & 9.76\%  & 0.56   & \multicolumn{1}{c|}{0.34} & 97.43\% & 12.05\% & 0.52   & \multicolumn{1}{c|}{0.98} & 93.77\% \\
EP-Diff        & 12.20\% & 0.38   & \multicolumn{1}{c|}{1.39} & 96.83\% & 13.84\% & 0.45   & \multicolumn{1}{c|}{0.77} & 89.44\% \\ \hline
               & \multicolumn{4}{c|}{JPN-IDN(1975$\sim$1995)}                & \multicolumn{4}{c|}{JPN-IDN(1995$\sim$2020)}                \\ \cline{2-9} 
               & MR      & ST-MSE & \multicolumn{1}{c|}{$\Delta$mag}  & FAR     & MR      & ST-MSE & \multicolumn{1}{c|}{$\Delta$mag}  & FAR     \\ \hline
ETAS           & 97.42\% & 0.12   & \multicolumn{1}{c|}{0.51} & 97.89\% & 97.85\% & 0.10   & \multicolumn{1}{c|}{0.64} & 98.34\% \\
EP-CNN         & 2.09\%  & 0.59   & \multicolumn{1}{c|}{0.44} & 78.46\% & 8.99\%  & 0.53   & \multicolumn{1}{c|}{0.72} & 95.77\% \\
EP-LSTM        & 4.83\%  & 0.74   & \multicolumn{1}{c|}{0.69} & 93.67\% & 9.91\%  & 0.43   & \multicolumn{1}{c|}{0.86} & 92.53\% \\
EP-Transformer & 8.21\%  & 0.55   & \multicolumn{1}{c|}{0.97} & 96.65\% & 12.50\% & 0.56   & \multicolumn{1}{c|}{0.67} & 94.26\% \\
EP-Diff        & 9.98\%  & 0.45   & \multicolumn{1}{c|}{0.56} & 95.98\% & 15.70\% & 0.41   & \multicolumn{1}{c|}{0.63} & 91.69\% \\ \hline
               & \multicolumn{4}{c|}{AU South(1975$\sim$1995)}               & \multicolumn{4}{c|}{AU South(1995$\sim$2020)}               \\ \cline{2-9} 
               & MR      & ST-MSE & \multicolumn{1}{c|}{$\Delta$mag}  & FAR     & MR      & ST-MSE & \multicolumn{1}{c|}{$\Delta$mag}  & FAR     \\ \hline
ETAS           & 76.32\% & 0.17   & \multicolumn{1}{c|}{1.19} & 99.52\% & 74.78\% & 0.17   & \multicolumn{1}{c|}{1.08} & 99.61\% \\
EP-CNN         & 0\%     & -      & \multicolumn{1}{c|}{-}    & 100\%   & 0\%     & -      & \multicolumn{1}{c|}{-}    & 100\%   \\
EP-LSTM        & 0\%     & -   & \multicolumn{1}{c|}{-} & 97.91\% & 4.23\%  & 0.72   & \multicolumn{1}{c|}{0.36} & 93.86\% \\
EP-Transformer & 2.63\%  & 0.51   & \multicolumn{1}{c|}{0.46} & 96.42\% & 5.63\%  & 0.75   & \multicolumn{1}{c|}{0.23} & 92.05\% \\
EP-Diff        & 7.89\%  & 0.42   & \multicolumn{1}{c|}{0.43} & 94.75\% & 11.27\% & 0.62   & \multicolumn{1}{c|}{0.14} & 92.03\% \\ \hline
               & \multicolumn{4}{c|}{USA-CL(1975$\sim$1995)}                 & \multicolumn{4}{c|}{USA-CL(1995$\sim$2020)}                 \\ \cline{2-9} 
               & MR      & ST-MSE & \multicolumn{1}{c|}{$\Delta$mag}  & FAR     & MR      & ST-MSE & \multicolumn{1}{c|}{$\Delta$mag}  & FAR     \\ \hline
ETAS           & 98.12\% & 0.10   & \multicolumn{1}{c|}{0.49} & 99.73\% & 98.68\% & 0.09   & \multicolumn{1}{c|}{0.58} & 99.81\% \\
EP-CNN         & 2.22\%  & 0.87   & \multicolumn{1}{c|}{0.64} & 81.25\% & 7.89\%  & 0.79   & \multicolumn{1}{c|}{0.92} & 88.59\% \\
EP-LSTM        & 4.13\%  & 0.94   & \multicolumn{1}{c|}{0.66} & 96.73\% & 13.85\% & 0.37   & \multicolumn{1}{c|}{0.47} & 93.21\% \\
EP-Transformer & 5.71\%  & 0.45   & \multicolumn{1}{c|}{0.50} & 97.78\% & 16.59\% & 0.34   & \multicolumn{1}{c|}{0.45} & 91.38\% \\
EP-Diff        & 8.25\%  & 0.44   & \multicolumn{1}{c|}{0.67} & 96.09\% & 24.32\% & 0.29   & \multicolumn{1}{c|}{0.81} & 87.27\% \\ \hline
               & \multicolumn{4}{c|}{Atlantic(1975$\sim$1995)}               & \multicolumn{4}{c|}{Atlantic(1995$\sim$2020)}               \\ \cline{2-9} 
               & MR      & ST-MSE & \multicolumn{1}{c|}{$\Delta$mag}  & FAR     & MR      & ST-MSE & \multicolumn{1}{c|}{$\Delta$mag}  & FAR     \\ \hline
ETAS           & 92.68\% & 0.14   & \multicolumn{1}{c|}{1.69} & 98.95\%  & 89.62\% & 0.19   & \multicolumn{1}{c|}{1.45} & 98.96\% \\
EP-CNN         & 0\%     & -   & \multicolumn{1}{c|}{-} & 91.24\% & 4.57\%  & 0.46   & \multicolumn{1}{c|}{0.39} & 93.68\% \\
EP-LSTM        & 2.44\%  & 0.73   & \multicolumn{1}{c|}{0.82} & 93.24\% & 10.66\% & 0.78   & \multicolumn{1}{c|}{0.99} & 92.36\% \\
EP-Transformer & 2.44\%  & 0.41   & \multicolumn{1}{c|}{0.79} & 92.65\% & 15.74\% & 0.26   & \multicolumn{1}{c|}{1.07} & 90.49\% \\
EP-Diff        & 7.32\%  & 0.37   & \multicolumn{1}{c|}{0.54} & 95.42\% & 24.37\% & 0.17   & \multicolumn{1}{c|}{1.42} & 91.92\% \\ \hline
               & \multicolumn{4}{c|}{AfricaAsia(1975$\sim$1995)}             & \multicolumn{4}{c|}{AfricaAsia(1995$\sim$2020)}             \\ \cline{2-9} 
               & MR      & ST-MSE & \multicolumn{1}{c|}{$\Delta$mag}  & FAR     & MR      & ST-MSE & \multicolumn{1}{c|}{$\Delta$mag}  & FAR     \\ \hline
ETAS           & 100\%   & 0.28   & \multicolumn{1}{c|}{0.86} & 98.80\% & 100\%   & 0.24   & \multicolumn{1}{c|}{0.91} & 99.11\% \\
EP-CNN         & 0\%     & -      & \multicolumn{1}{c|}{-}    & 100\%   & 0\%     & -      & \multicolumn{1}{c|}{-}    & 100\%   \\
EP-LSTM        & 0\%     & -   & \multicolumn{1}{c|}{-} & 91.85\% & 0\%     & -      & \multicolumn{1}{c|}{-}    & 100\%   \\
EP-Transformer & 0\%     & -   & \multicolumn{1}{c|}{-} & 96.52\% & 6.67\%  & 0.55   & \multicolumn{1}{c|}{0.74} & 98.06\% \\
EP-Diff        & 20\%    & 0.42   & \multicolumn{1}{c|}{0.73} & 89.44\% & 33.33\% & 0.27   & \multicolumn{1}{c|}{0.63} & 90.03\% \\ \hline
\end{tabular}}
\label{Table 2}
\end{table}

\begin{table}[]
\centering
\caption{Results of our control experiments regarding the moment tensor information. Note that ‘-’ indicates training using only basic information, while ‘+’ indicates the inclusion of additional moment tensor information. The implementation details can be found in the appendixes.}
 \scalebox{0.7}[0.8]{
\begin{tabular}{|l|cccc|cccc|}
\hline
               & \multicolumn{4}{c|}{-}                                & \multicolumn{4}{c|}{+}                                 \\ \cline{2-9} 
               & \multicolumn{4}{c|}{JPN-IDN(1996$\sim$2021)}               & \multicolumn{4}{c|}{JPN-IDN(1996$\sim$2021)}                \\ \cline{2-9} 
               & MR     & ST-MSE & \multicolumn{1}{c|}{$\Delta$mag}  & FAR     & MR      & ST-MSE & \multicolumn{1}{c|}{$\Delta$mag}  & FAR     \\ \hline
EP-LSTM        & 0\%    & -      & \multicolumn{1}{c|}{-}    & 100\%   & 9.62\%  & 0.39   & \multicolumn{1}{c|}{1.18} & 98.04\% \\
EP-Transformer & 0\%    & -      & \multicolumn{1}{c|}{-}    & 100\%   & 13.46\% & 0.34   & \multicolumn{1}{c|}{1.32} & 97.53\% \\
EP-Diff        & 3.85\% & 0.57   & \multicolumn{1}{c|}{0.86} & 98.22\% & 25.00\% & 0.28   & \multicolumn{1}{c|}{1.40} & 96.95\% \\ \hline
               & \multicolumn{4}{c|}{USA-CL(1996$\sim$2021))}               & \multicolumn{4}{c|}{USA-CL(1996$\sim$2021)}                 \\ \cline{2-9} 
               & MR     & ST-MSE & \multicolumn{1}{c|}{$\Delta$mag}  & FAR     & MR      & ST-MSE & \multicolumn{1}{c|}{$\Delta$ mag}  & FAR     \\ \hline
EP-LSTM        & 0\%    & -      & \multicolumn{1}{c|}{-}    & 100\%   & 12.90\% & 0.34   & \multicolumn{1}{c|}{1.29} & 94.59\% \\
EP-Transformer & 4.84\% & 0.39   & \multicolumn{1}{c|}{0.75} & 97.47\% & 17.74\% & 0.27   & \multicolumn{1}{c|}{1.37} & 98.26\% \\
EP-Diff        & 4.84\% & 0.45   & \multicolumn{1}{c|}{0.71} & 96.74\% & 24.19\% & 0.22   & \multicolumn{1}{c|}{1.31} & 98.61\% \\ \hline
\end{tabular}}
\label{Table 3}
\end{table}

\section{Limitations}
Overall, EPBench and the short-term earthquake prediction task still have the following limitations:
1. Other physical information related to earthquakes, such as geoelectric variations, geomagnetic variations, and crustal stress situations, could be incorporated into the dataset in the future. 2. We have not yet utilized waveform data, and the multimodal data has not yet been combined with the non-multimodal data. 3. Furthermore, We  need more specialized neural network architectures to better leverage earthquake-related information and, consequently, improve the models' performance.

\section{Conclusion}
We introduce EPBench, which, to our knowledge, is the first global regional-scale short-term earthquake prediction benchmark. We contributed (1) datasets with 924,472 earthquake records and 2959 multimodal earthquake records, (2) a set of intuitive and effective metrics tailored for the earthquake prediction task,(3)preliminary regional partitioning and  moment tensor information preprocessing with experiments demonstrating their effectiveness. We believe this benchmark will serve as a guide to attract more researchers to explore new methods for addressing this task, which holds great significance for human existence.

\medskip

{\small
\nocite{*}
\bibliographystyle{ieee_fullname}
\bibliography{neurips_2025}
}

\clearpage

\appendix

\section{Causes of Earthquakes}
\textit{\textbf{Knowledge from "https://www.drishtiias.com/to-the-points/paper1/earthquake-4".}}

An earthquake in simple words is the shaking of the earth. It is a natural event. It is caused due to release of energy, which generates waves that travel in all directions. The vibrations called seismic waves are generated from earthquakes that travel through the Earth and are recorded on instruments called seismographs.
The location below the earth’s surface where the earthquake starts is called the hypocenter, and the location directly above it on the surface of the earth is called the epicenter.

\textbf{Types of Earthquake and Causes}

Fault Zones: The release of energy occurs along a fault. A fault is a sharp break in the crustal rocks. Rocks along a fault tend to move in opposite directions. As the overlying rock strata press them, the friction locks them together.
However, their tendency to move apart at some point of time overcomes the friction. As a result, the blocks get deformed and eventually, they slide past one another abruptly.
This causes earthquake in the form of release of energy, and the energy waves travel in all directions.

\begin{figure}[h]
    \centering
    \includegraphics[width=0.8\linewidth,height=6.2cm]{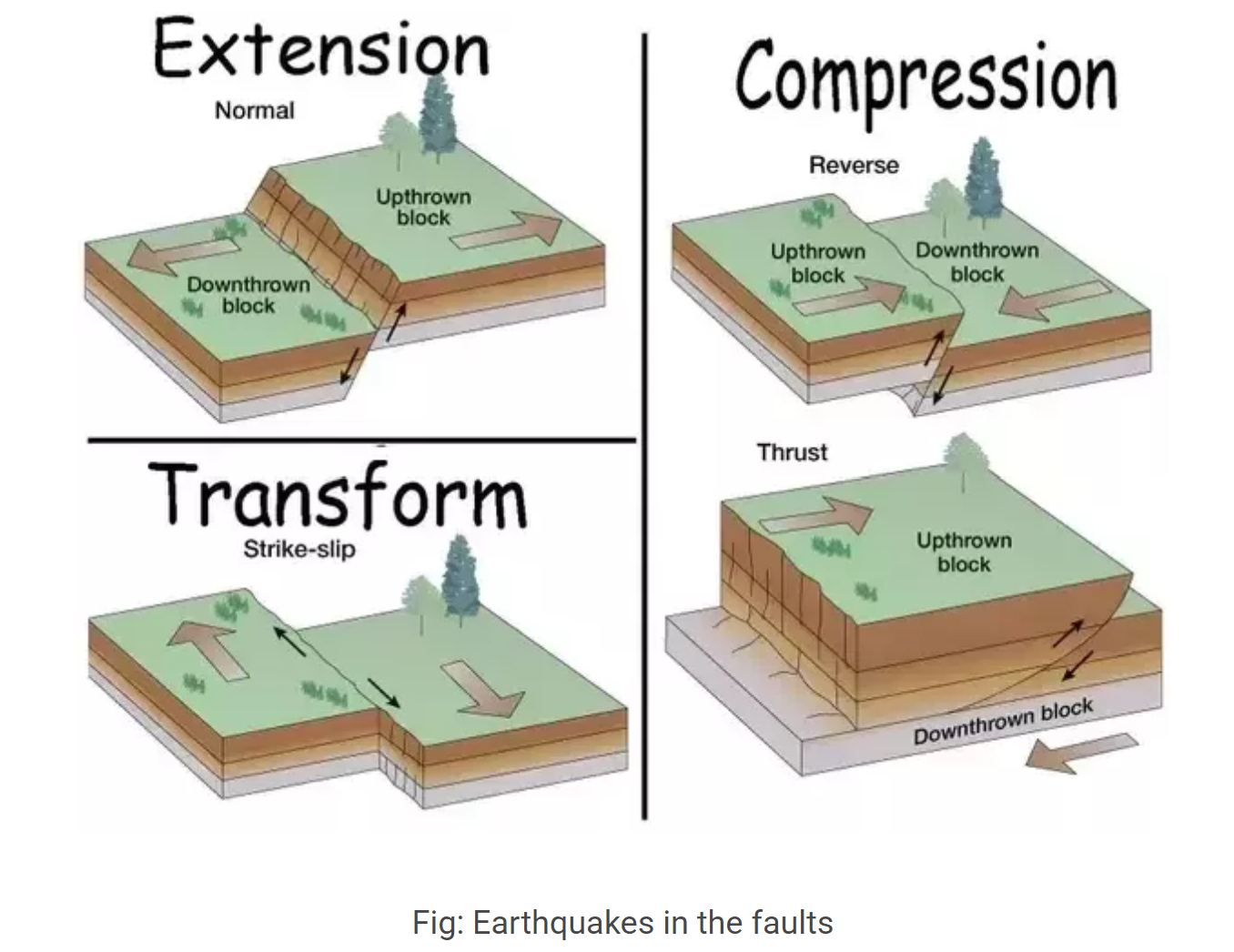}

\end{figure}

\textbf{1.Tectonic Earthquakes:}

The most common ones are the tectonic earthquakes.
Although the Earth looks like a pretty solid place from the surface, it’s actually extremely active just below the surface.
The Earth is made of four basic layers (generally three): a solid crust, a hot, nearly solid mantle, a liquid outer core and a solid inner core.

\begin{figure}[h]
    \centering
    \includegraphics[width=0.75\linewidth,height=6.2cm]{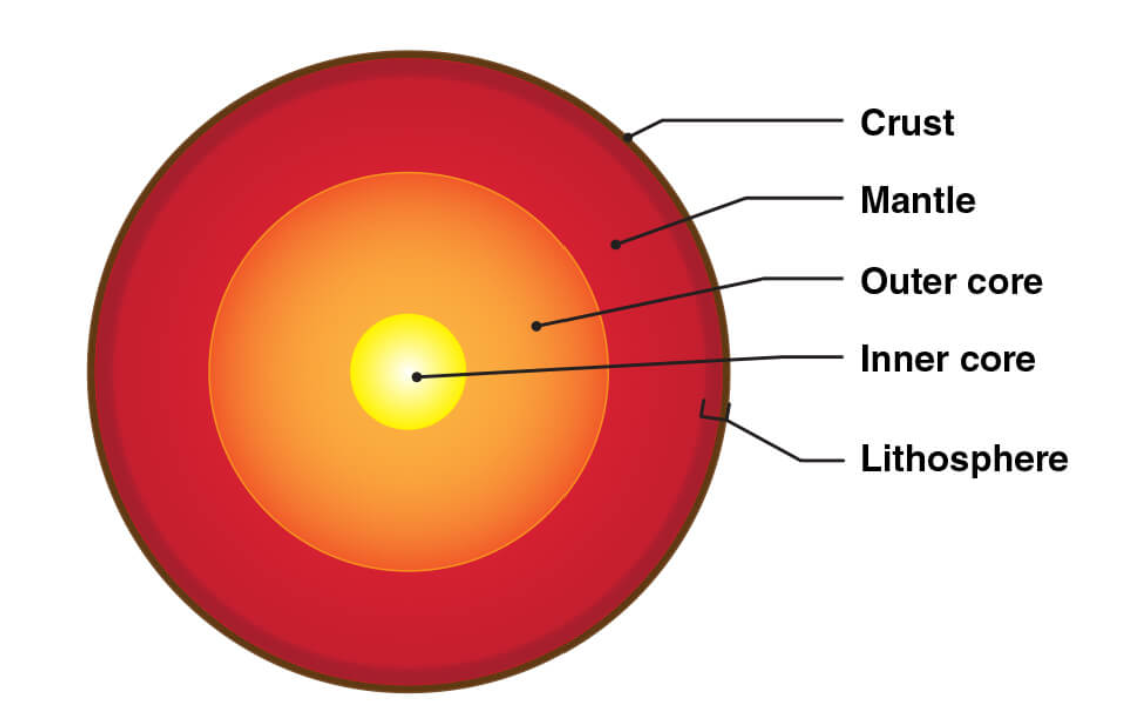}

\end{figure}
Tectonic plates (Lithospheric plates) are constantly shifting as they drift around on the viscous, or slowly flowing, mantle layer below.
Tectonic plates (Lithospheric plates) are constantly shifting as they drift around on the viscous, or slowly flowing, mantle layer below.
\begin{figure}[h]
    \centering
    \includegraphics[width=1.0\linewidth,height=7cm]{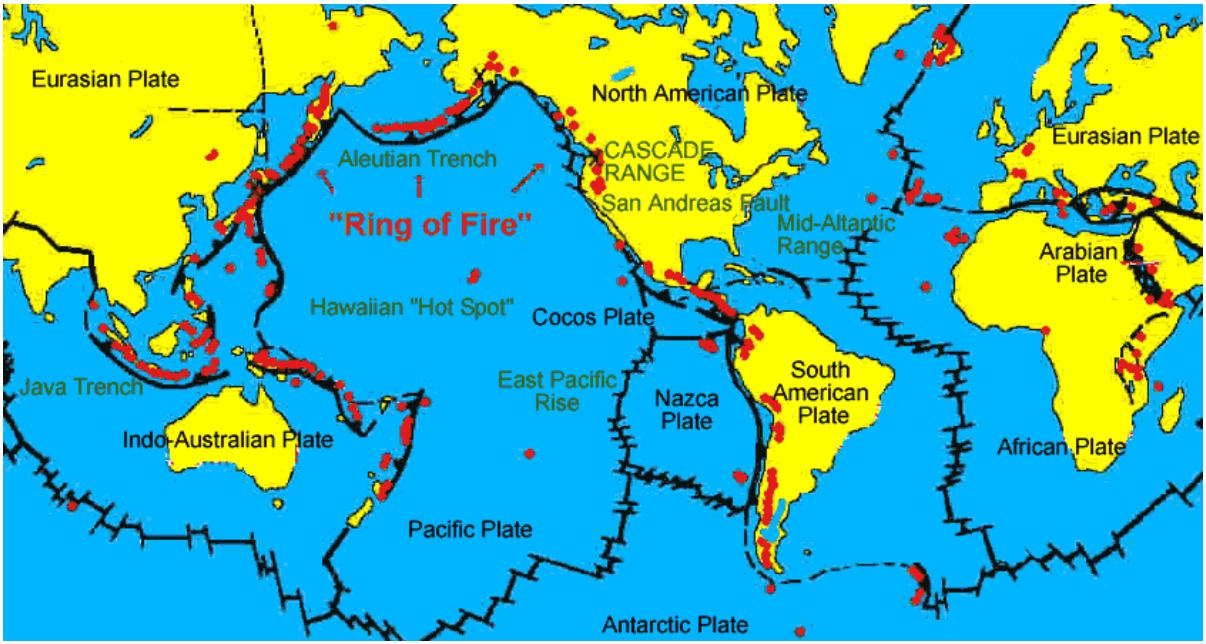}

\end{figure}

This non-stop movement causes stress on Earth’s crust. When the stresses get too large, it leads to cracks called faults.
When tectonic plates move, it also causes movements at the faults. Thus, the slipping of land along the faultline along convergent, divergent and transform boundaries cause earthquakes.
The point where the energy is released is called the focus of an earthquake, alternatively, it is called the hypocentre. The energy waves travelling in different directions reach the surface.
The point on the surface, nearest to the focus, is called epicentre. It is the first one to experience the waves. It is a point directly above the focus.

\textbf{2.Volcanic Earthquake}

A special class of tectonic earthquake is sometimes recognised as volcanic earthquake. However, these are confined to areas of active volcanoes.
Earthquakes produced by stress changes in solid rock due to the injection or withdrawal of magma (molten rock) are called volcano earthquakes.
These earthquakes can cause land to subside and can produce large ground cracks. These earthquakes can occur as rock is moving to fill in spaces where magma is no longer present.
Volcano-tectonic earthquakes don't indicate that the volcano will be erupting but can occur at any time.

\textbf{3.Human Induced Earthquakes}

In the areas of intense mining activity, sometimes the roofs of underground mines collapse causing minor tremors. These are called collapse earthquakes.
Ground shaking may also occur due to the explosion of chemical or nuclear devices. Such tremors are called explosion earthquakes.
The earthquakes that occur in the areas of large reservoirs are referred to as reservoir induced earthquakes.

\section{Moment Tensor}
\begin{figure}[h]
    \centering
    \includegraphics[width=0.7\linewidth,height=7cm]{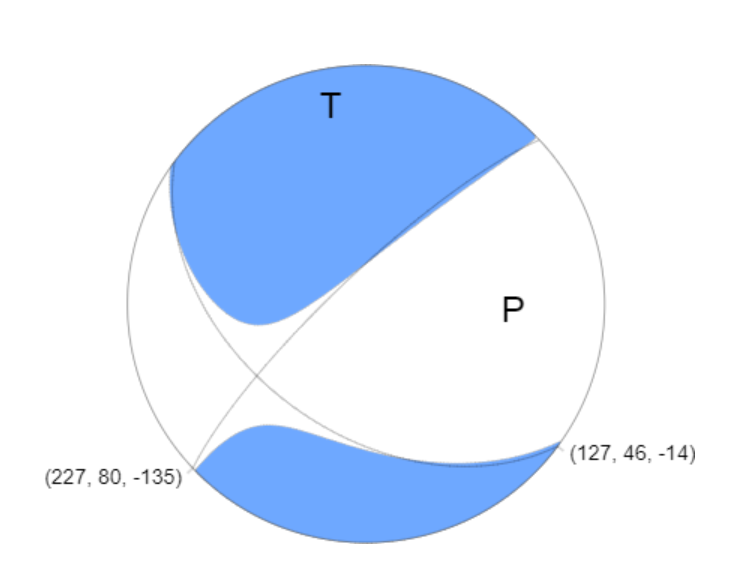}

\end{figure}
\begin{figure}[h]
    \centering
    \includegraphics[width=0.82\linewidth,height=6.6cm]{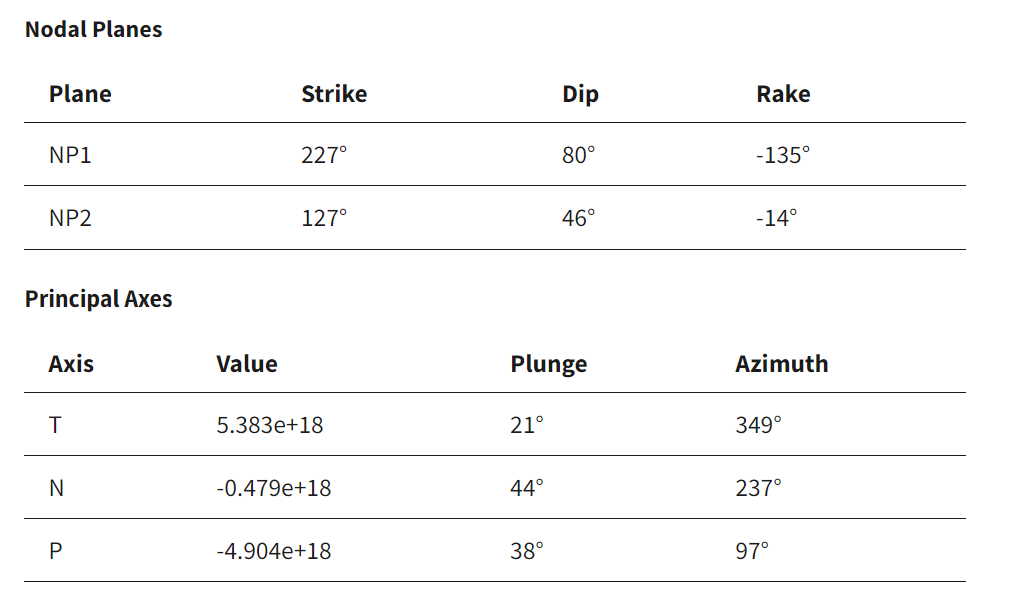}
  \caption{An example of recorded moment tensor information.}
   
\end{figure}
\subsection{Nodal Planes}
\textbf{Definition} During an earthquake, the radiation pattern generated by fault slip contains two mutually perpendicular planes (NP1 and NP2), known as nodal planes. One corresponds to the actual fault plane, while the other is the auxiliary plane.

\textbf{Angular Parameters}
\begin{itemize}
    \item Strike: The angle between the nodal plane and geographic north (0°–360°).
    \item Dip: The inclination of the nodal plane relative to the horizontal plane (0°–90°).
    \item Rake: The angle between the projection of the fault slip direction onto the nodal plane and the strike direction (-180°–180°).
\end{itemize}

\textbf{Significance} The parameters of NP1 and NP2 fully describe the fault geometry and slip direction, but the actual fault plane cannot be distinguished from the moment tensor alone.

\subsection{Axes}
\textbf{Definition} Three orthogonal principal axes derived from the eigenvalue decomposition of the moment tensor, corresponding to principal stress directions:
\begin{itemize}
    \item T-axis (Tension Axis): Direction of the largest positive eigenvalue, representing the maximum extensional (tensional) direction.

\item P-axis (Pressure Axis): Direction of the largest negative eigenvalue, representing the maximum compressional direction.

\item N-axis (Neutral Axis): Direction of the intermediate eigenvalue, typically near zero (strictly zero in a double-couple model).
\end{itemize}

\textbf{Angular Parameters}
\begin{itemize}
    \item Azimuth: Angle between the horizontal projection of the axis and geographic north.
    \item Plunge: Angle between the axis and the horizontal plane (positive downward).
\end{itemize}
\textbf{Value}  The "values" associated with the T, N, and P axes are the eigenvalues of the moment tensor, with units of moment (N·m). They reflect the strength of the force couples in each direction, where the sign indicates the force type (positive for extension, negative for compression).

\subsection{Summary}
\textbf{NP1 and NP2}: Describe potential fault planes using strike, dip, and slip angles.

\textbf{T, N, P Axes}: Principal axes derived from moment tensor decomposition; directions are defined by eigenvectors, and eigenvalues represent moment magnitude and force type (extension/compression).

\textbf{Angles}: Describe nodal plane geometry (strike, dip, slip) or principal axis orientation (azimuth, plunge).

This framework provides the mathematical and geometric basis for analyzing focal mechanisms (e.g., fault type, stress field orientation).

\section{Proof of Haversine Formula}
\begin{figure}[h]
    \centering
    \includegraphics[width=00.5\linewidth,height=5.2cm]{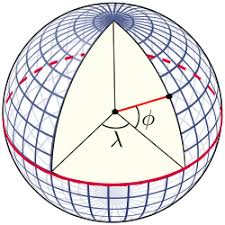}
    \caption{Spherical Coordinate System}
   
\end{figure}

\textbf{1. Spherical to Cartesian Coordinates Conversion}

Assume Earth is a perfect sphere with radius 
$R$.

Convert two points $P_1(\phi_1, \lambda_1)$ and $P_2(\phi_2, \lambda_2)$ to 3D Cartesian coordinates:
\[
\begin{cases}
x = R \cos \phi \cos \lambda,\\
y = R \cos \phi \sin \lambda,\\
z = R \sin \phi,
\end{cases}
\]
where:
\begin{itemize}
  \item $\phi$: Latitude (positive for north, negative for south),
  \item $\lambda$: Longitude (positive for east, negative for west).
\end{itemize}

\textbf{2.Central Angle $\theta$ Between Two Points}

The central angle $\theta$ subtended by the two points at the sphere's center is derived using the dot product:
\[
\cos \theta = \frac{\mathbf{P}_1 \cdot \mathbf{P}_2}{R^2}.
\]
Expanding the dot product:
\[
\mathbf{P}_1 \cdot \mathbf{P}_2 = R^2 [\cos \phi_1 \cos \phi_2 \cos (\Delta \lambda) + \sin \phi_1 \sin \phi_2],
\]
where $\Delta \lambda = \lambda_2 - \lambda_1$. Thus:
\[
\cos \theta = \cos \phi_1 \cos \phi_2 \cos (\Delta \lambda) + \sin \phi_1 \sin \phi_2.
\]

\textbf{3.Derivation of the Haversine Formula}

To avoid numerical instability for small angles, use the \textbf{half-angle identity} and define the \textbf{Haversine function}:
\[
{hav}(\theta) = \sin^2 \left( \frac{\theta}{2} \right).
\]
From the identity $\cos \theta = 1 - 2 \sin^2 \left( \frac{\theta}{2} \right)$, we get:
\[
\sin^2 \left( \frac{\theta}{2} \right) = \frac{1 - \cos \theta}{2}.
\]
Substitute $\cos \theta$ from Step 2:
\[
{hav}(\theta) = \frac{1 - [\cos \phi_1 \cos \phi_2 \cos (\Delta \lambda) + \sin \phi_1 \sin \phi_2]}{2}.
\]
Split the numerator:
\[
1 - \cos \phi_1 \cos \phi_2 \cos (\Delta \lambda) - \sin \phi_1 \sin \phi_2.
\]
Using the identity $\sin \phi_1 \sin \phi_2 = \frac{1}{2} [\cos (\phi_1 - \phi_2) - \cos (\phi_1 + \phi_2)]$, we can rearrange to obtain:
\[
{hav}(\theta) = {hav}(\Delta \phi) + \cos \phi_1 \cos \phi_2 {hav} (\Delta \lambda),
\]
where $\Delta \phi = \phi_2 - \phi_1$, and ${hav}(\Delta \phi) = \sin^2 \left(\frac{\Delta \phi}{2}\right)$.

\textbf{4.Final Distance Formula}

The central angle $\theta$ is:
\[
\theta = 2 \arcsin \left( \sqrt{{hav}(\theta)} \right).
\]
Substitute the Haversine expression:
\[
\theta = 2 \arcsin \left( \sqrt{ \sin^2 \left( \frac{\Delta \phi}{2} \right) + \cos \phi_1 \cos \phi_2 \sin^2 \left( \frac{\Delta \lambda}{2} \right) } \right).
\]
The great-circle distance $d$ becomes:
\[
d = R \cdot \theta = 2 R \arcsin \left( \sqrt{a} \right),
\]
where
\[
a = \sin^2 \left( \frac{\Delta \phi}{2} \right) + \cos \phi_1 \cos \phi_2 \sin^2 \left( \frac{\Delta \lambda}{2} \right).
\]

\newpage

\section{Dataset Analysis}
Oour dataset consists of two categories of data: one category includes all recorded earthquakes with magnitudes above 2.5 from 1970 to 2020 containing key information such as latitude, longitude, time, magnitude and depth; the other category includes  multimodal data with magnitudes above 6 from 1996 to 2021, which not only contains key information but also physical quantities such as waveform and  moment tensors. Our data is organized in CSV file format, with each column representing a type of information, such as time, latitude, and longitude. Each waveform-type data has a separate CSV file, corresponding to the ID number of the seismic event.

\subsection{Temporal Distribution Analysis}
\begin{figure}[h]
    \centering
    \includegraphics[width=1\linewidth,height=7.6cm]{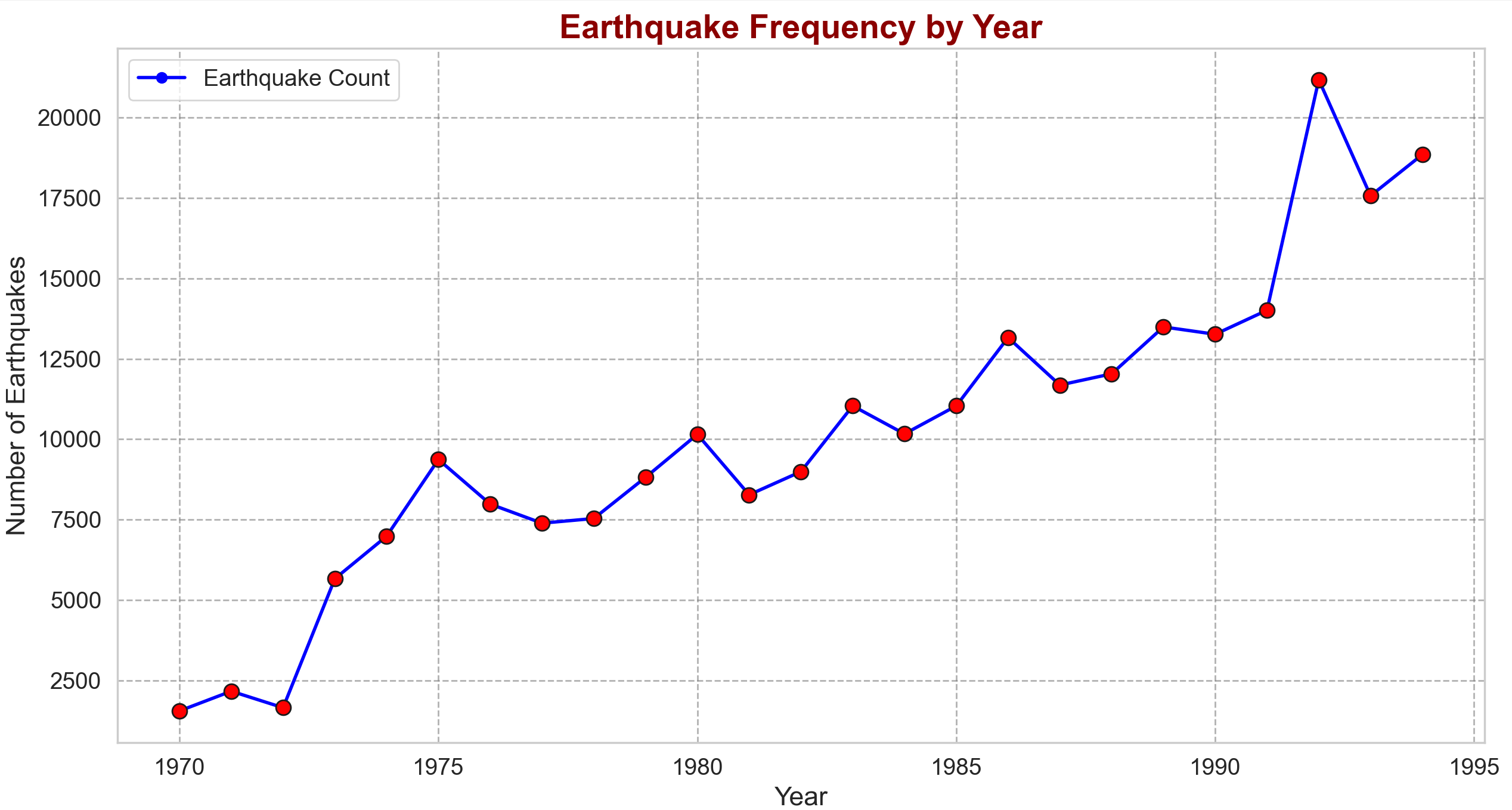}
    \caption{The annual frequency of earthquakes with magnitudes $\geq$2.5 from 1975 to 1994.}
   
\end{figure}
\begin{figure}[h]
    \centering
    \includegraphics[width=1\linewidth,height=7.6cm]{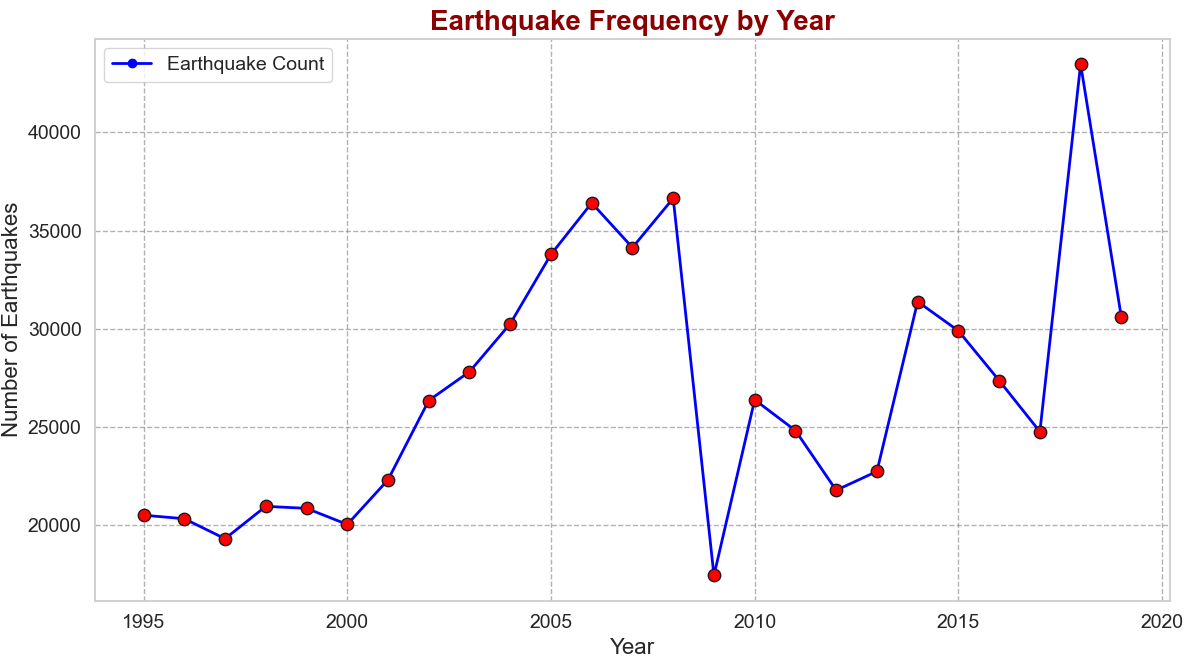}
    \caption{The annual frequency of earthquakes with magnitudes $\geq$2.5 from 1995 to 2019.}
   
\end{figure}

\subsection{Spatial Distribution Analysis}
\begin{figure}[h]
    \centering
    \includegraphics[width=1\linewidth,height=8.9cm]{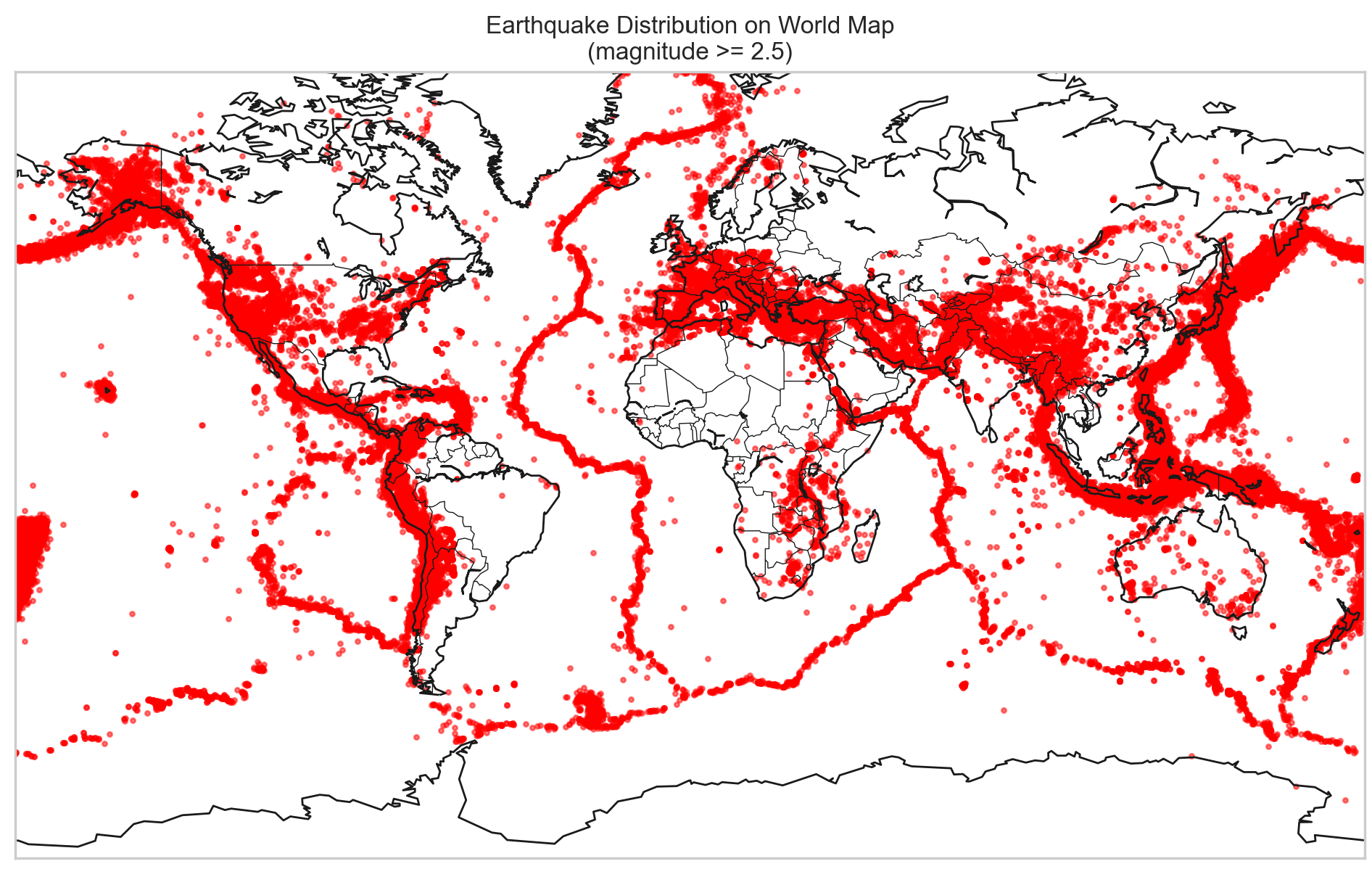}
    \caption{The spatial distribution of earthquakes with magnitudes $\geq$2.5 from 1970 to 1995}
   
\end{figure}
\begin{figure}[h]
    \centering
    \includegraphics[width=1\linewidth,height=8.9cm]{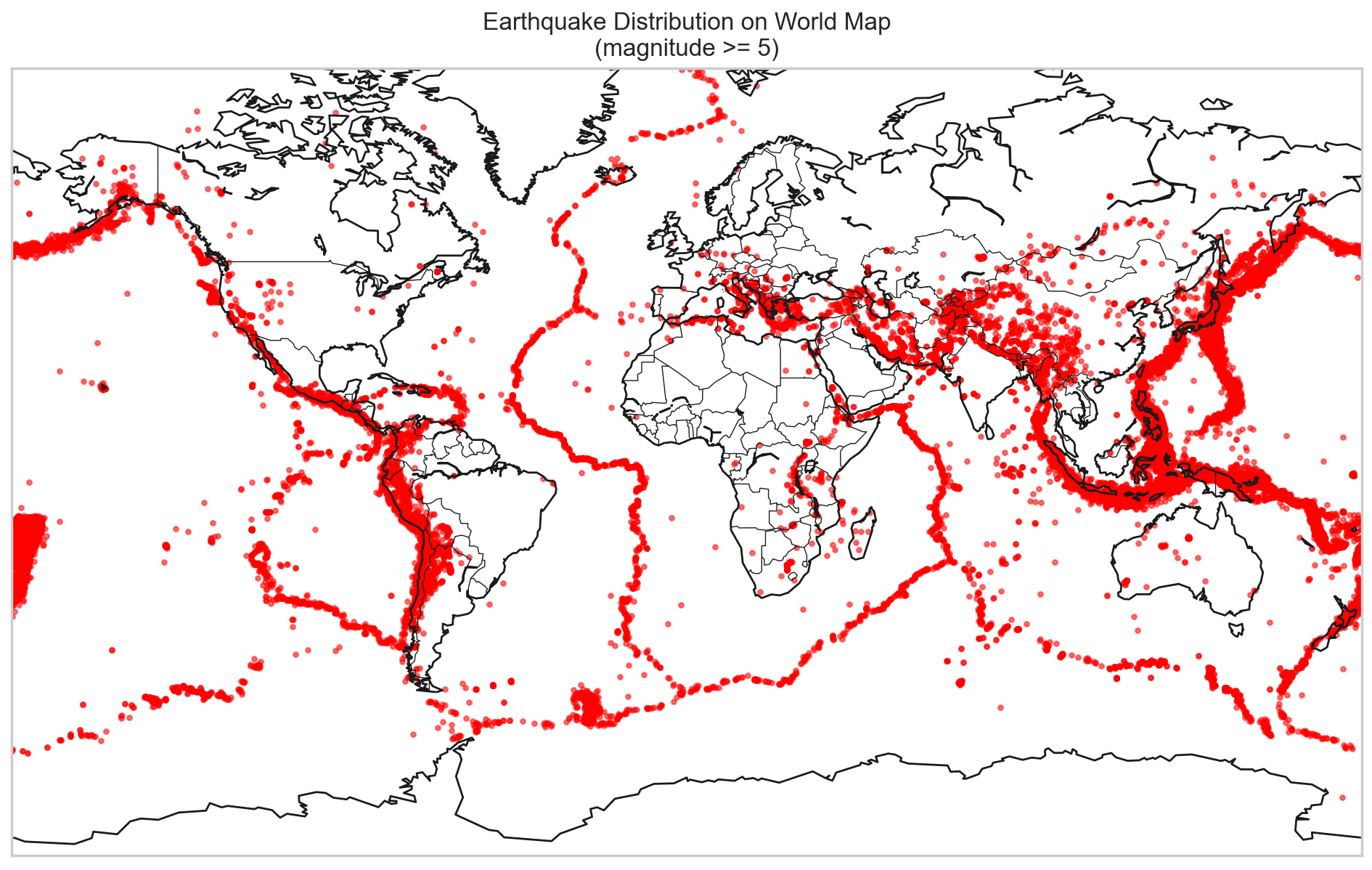}
    \caption{The spatial distribution of earthquakes with magnitudes $\geq$5 from 1970 to 1995}
   
\end{figure}
\begin{figure}[h]
    \centering
    \includegraphics[width=1.0\linewidth,height=8.9cm]{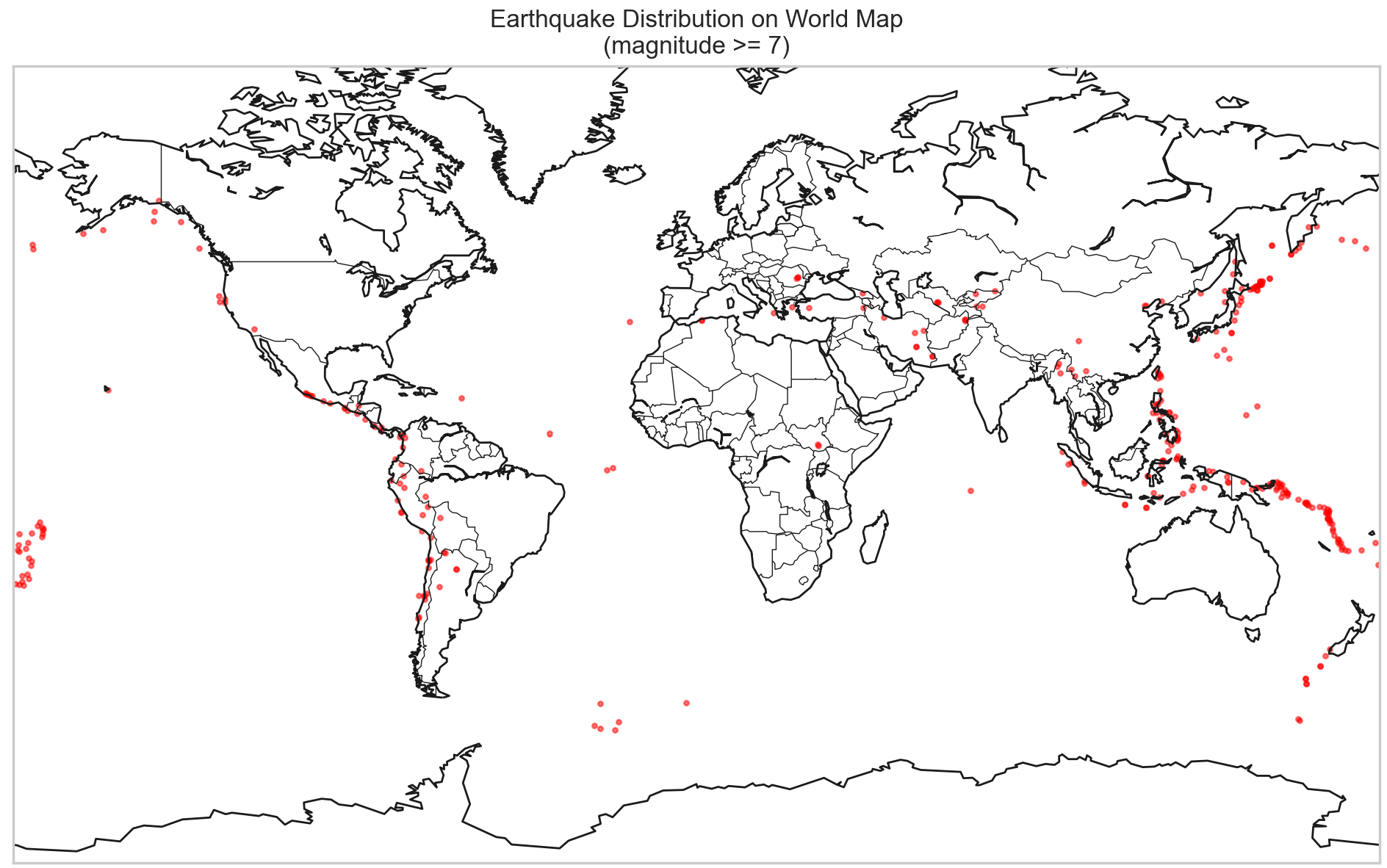}
    \caption{The spatial distribution of earthquakes with magnitudes $\geq$7 from 1970 to 1995}
   
\end{figure}

\begin{figure}[h]
    \centering
    \includegraphics[width=1\linewidth,height=8.9cm]{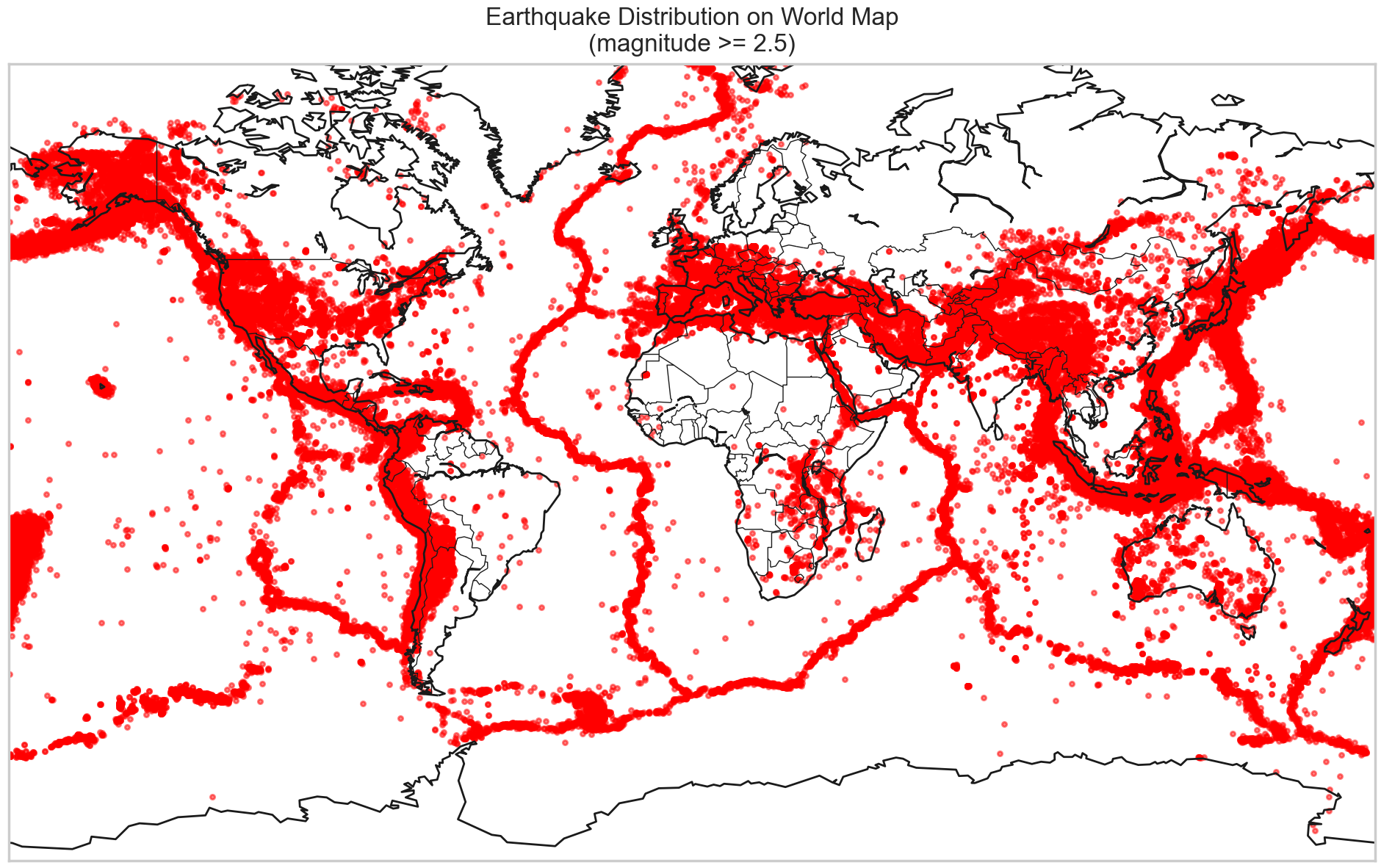}
    \caption{The spatial distribution of earthquakes with magnitudes $\geq$2.5 from 1995 to 2020}
   
\end{figure}
\begin{figure}[h]
    \centering
    \includegraphics[width=1\linewidth,height=8.9cm]{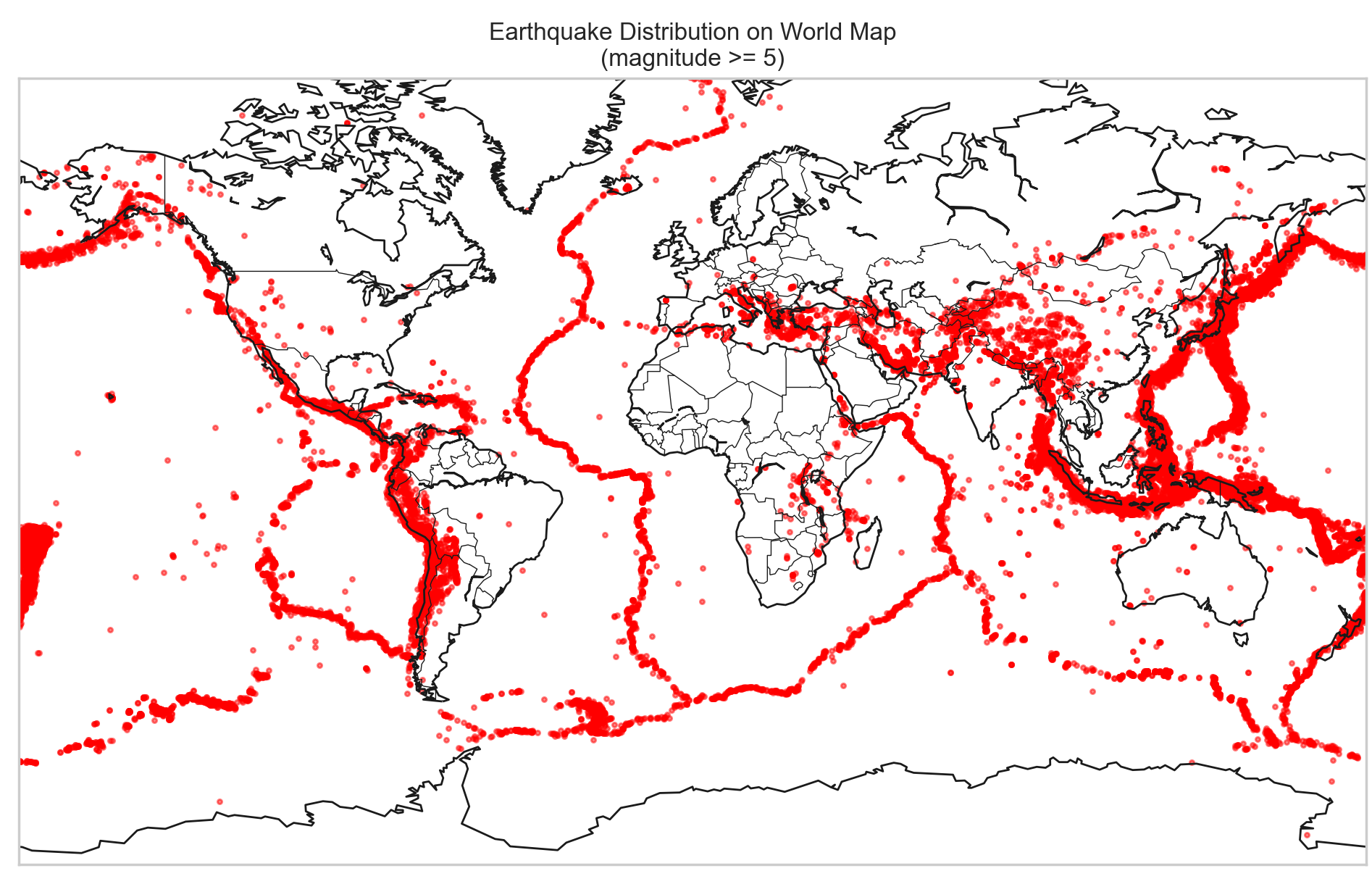}
    \caption{The spatial distribution of earthquakes with magnitudes $\geq$5 from 1995 to 2020}
   
\end{figure}
\begin{figure}[h]
    \centering
    \includegraphics[width=1.0\linewidth,height=8.9cm]{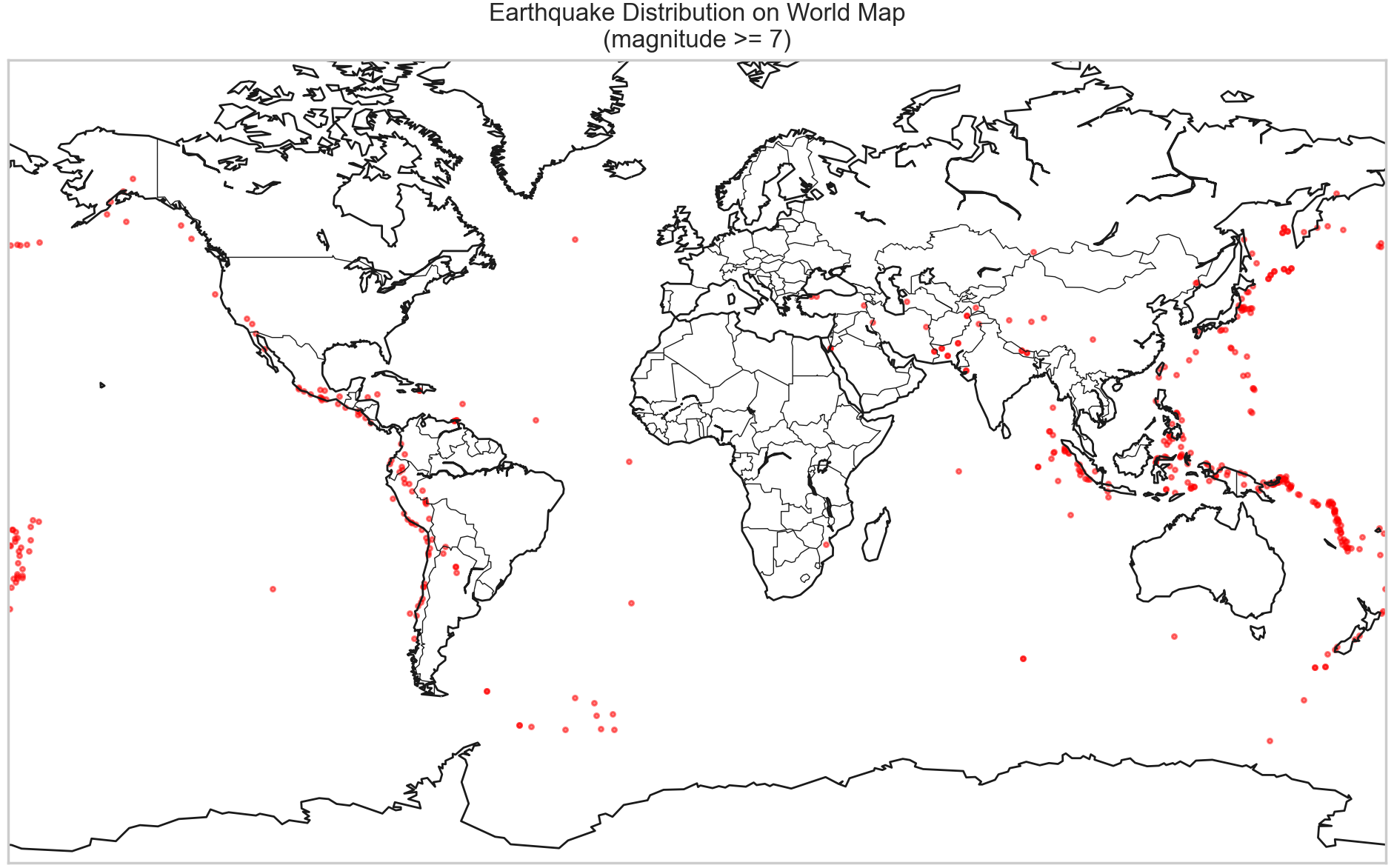}
    \caption{The spatial distribution of earthquakes with magnitudes $\geq$7 from 1995 to 2020}
   
\end{figure}

\clearpage

\subsection{Magnitude Distribution Analysis}

\begin{figure}[h]
    \centering
    \includegraphics[width=1\linewidth,height=8.9cm]{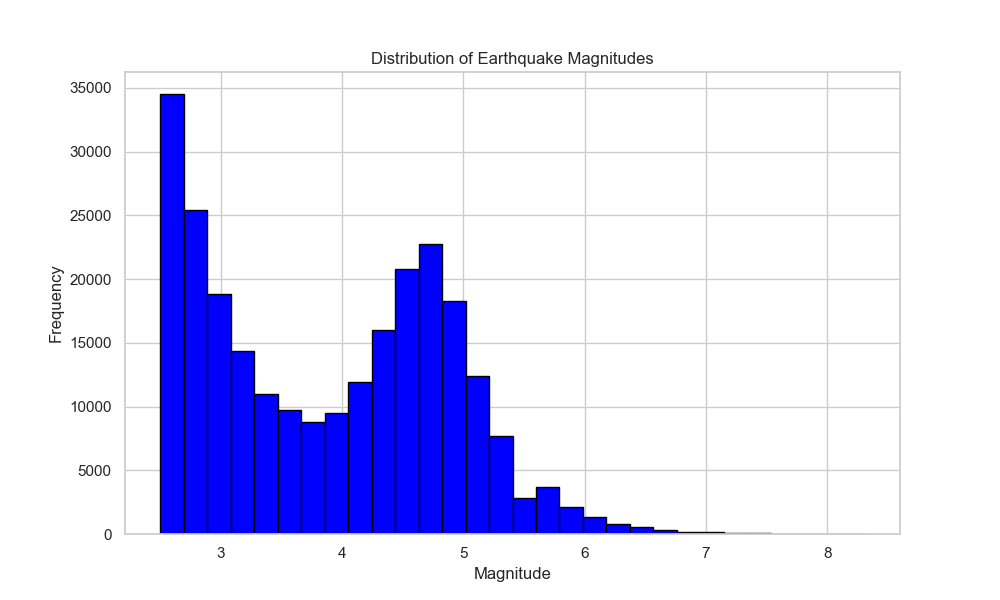}
    \caption{The magnitude distribution of earthquakes from 1970 to 1995}
   
\end{figure}
\begin{figure}[h]
    \centering
    \includegraphics[width=1.0\linewidth,height=8.9cm]{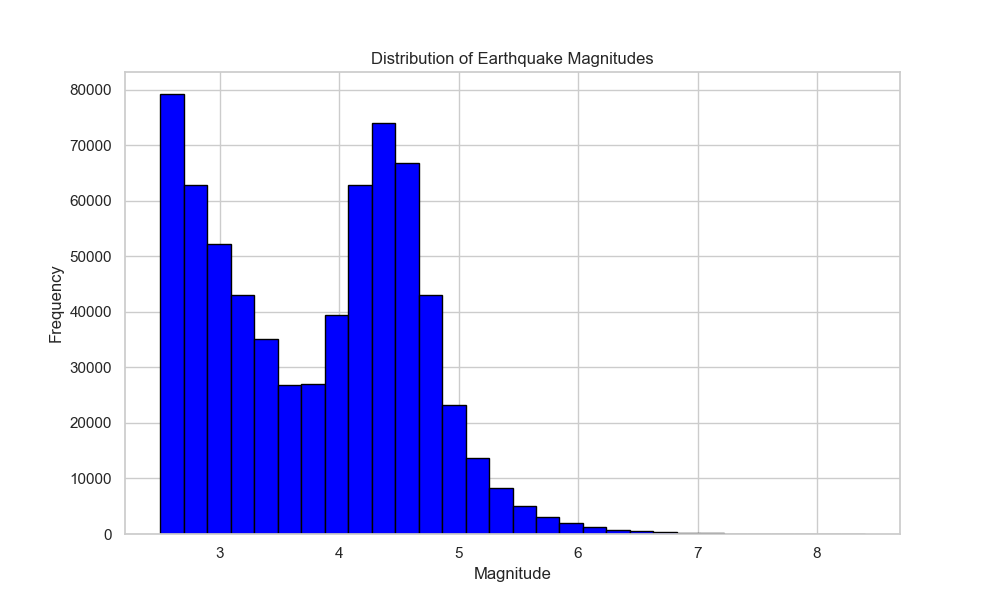}
    \caption{The magnitude distribution of earthquakes from 1995 to 2020}
   
\end{figure}

\clearpage

\subsection{Depth Distribution Analysis}
\begin{figure}[h]
    \centering
    \includegraphics[width=1\linewidth,height=8.9cm]{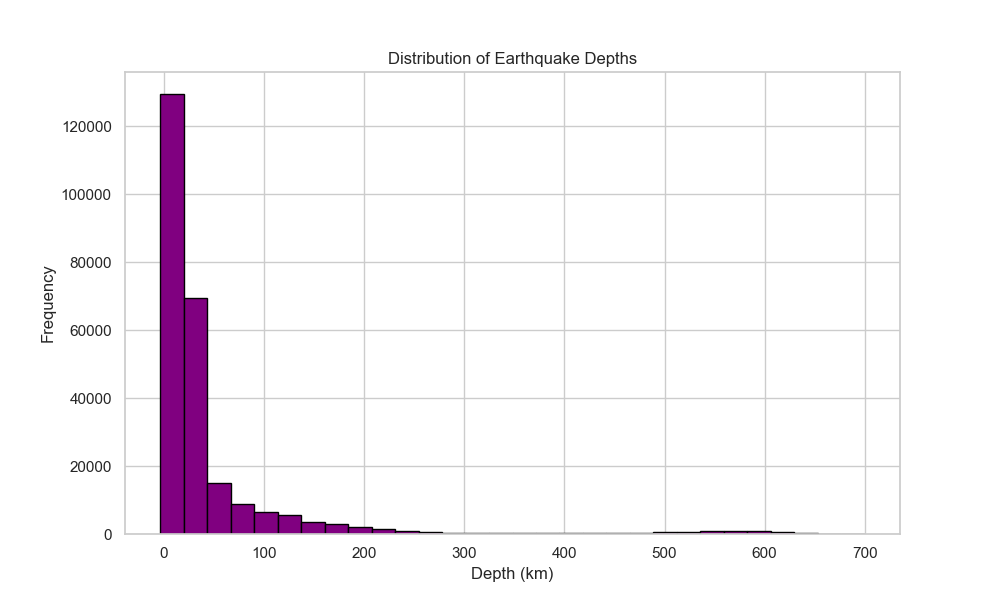}
    \caption{The depth distribution of earthquakes from 1970 to 1995}
   
\end{figure}
\begin{figure}[h]
    \centering
    \includegraphics[width=1.0\linewidth,height=8.9cm]{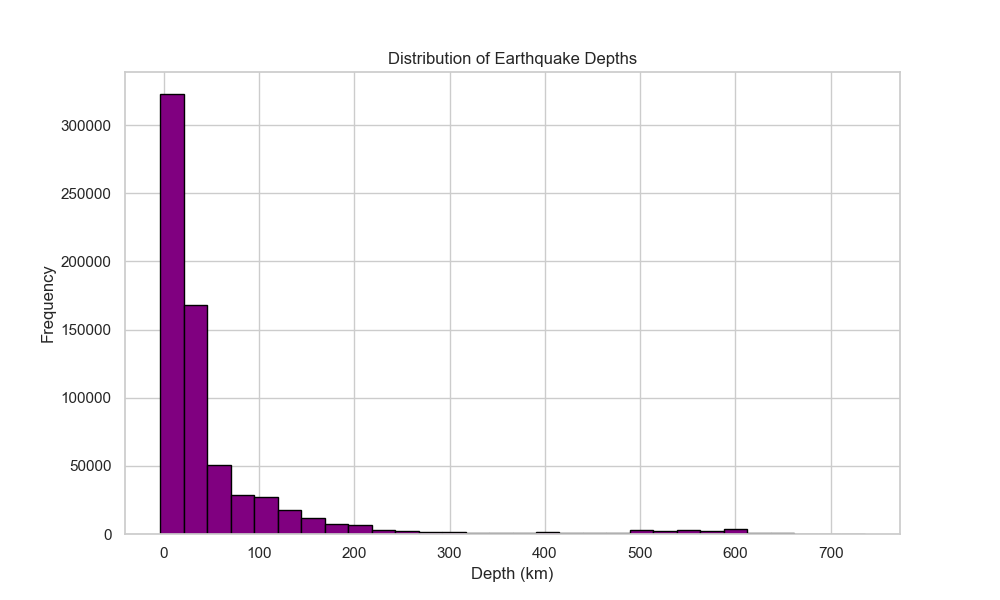}
    \caption{The depth distribution of earthquakes from 1995 to 2020}
   
\end{figure}
\clearpage
\subsection{Magnitude-Depth Pair Distribution }
\begin{figure}[h]
    \centering
    \includegraphics[width=1.1\linewidth,height=9.2cm]{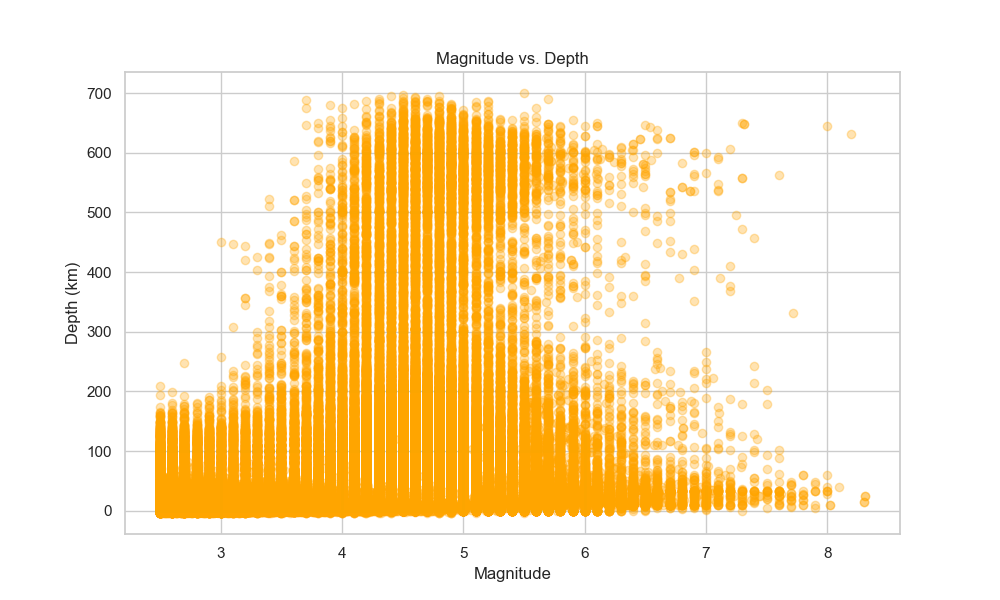}
    \caption{The magnitude-depth pair distribution of earthquakes.}
   
\end{figure}

\section{Model Architecture}
\subsection{EP-CNN}
The proposed EP-CNN  employs a hierarchical 1D convolutional architecture designed for temporal feature extraction and regression tasks. The model consists of two principal components: a pyramidal feature extractor and a  regressor module.

\textbf{1. Pyramidal Feature Extractor}

Five consecutive convolution blocks form a contracting pyramidal structure:
\begin{itemize}
    \item Block 1: 512 filters (kernel=7, dilation=5) with padding=15
    \item Block 2: 256 filters (kernel=5, dilation=4) with padding=8
    \item Block 3: 128 filters (kernel=5, dilation=3) with padding=6
    \item Block 4: 128 filters (kernel=3, dilation=2) with padding=2
    \item Block 5: 128 filters (kernel=3) with padding=2 
\end{itemize}

Each convolutional layer incorporates batch normalization and ReLU activation followed by  pooling operations. Dropout regularization after each pooling operation mitigates overfitting.

\textbf{2. Regression Module}

The regression head comprises two fully-connected layers (128 hidden units) with ReLU non-linearity, mapping the compressed temporal representation to target output.

\subsection{EP-LSTM}
The proposed EP-LSTM model implements a deep stacked LSTM architecture with progressive feature compression for sequential data modeling. The model employs four cascaded LSTM layers with pyramidal dimensionality reduction:
\begin{itemize}
    \item Layer 1: 256 hidden units → BatchNorm + Dropout(0.1)
    \item Layer 2: 128 hidden units → BatchNorm + Dropout(0.1)
    \item Layer 3: 64 hidden units → BatchNorm + Dropout(0.1)
    \item Layer 4: 64 hidden units → BatchNorm + Dropout(0.1)

\end{itemize}

Each LSTM layer preserves the original sequence length through full sequence output while progressively compressing feature dimensions. The conserved  resolution enables fine-grained temporal pattern preservation before final aggregation.

Following the LSTM layers, a linear layer first maps concatenated features from all time steps to a 64-dimensional latent space, subsequently activated by ReLU. A final linear projection layer then transforms these compressed representations to the target output.

\subsection{EP-Transformer}
The proposed EP-Transformer first employs a feature transformation module,which consists of three fully connected layers with GELU activation functions, progressively projecting the input dimension to 256 dimensions. 

Next, the model applies learnable positional encoding to incorporate positional information for each element in the sequence, enhancing the model's awareness of sequential order. The model then includes six Transformer encoder layers , each utilizing multi-head self-attention (8 heads) and a 512-dimensional feedforward network, enabling the model to capture complex dependencies within the sequence through layer-wise encoding.

After encoding, the model compresses the sequence dimension to 1 using adaptive average pooling . Finally, a classification head composed of fully connected layers, GELU activation, and dropout produces an output.

\subsection{EP-Diff}
The proposed EP-Diff consists of three core components: (1) A condition encoder built with a 4-layer Transformer (4 attention heads, 256-dim FFN) that processes conditional inputs; (2) A sinusoidal time embedding module that encodes diffusion timesteps via learned embeddings followed by a SiLU-activated projection; and (3) A noise prediction network implemented as an MLP with LayerNorm and SiLU activations (256→128→input\_dim) that combines the condition vector, noisy input, and time embedding.

The diffusion process follows standard DDPM methodology with linear scheduling ($\beta$=1e-4→0.02 over T=1000 steps). During generation, it progressively denoises random initialization through:
1.Condition-aware noise prediction via the trained model, 2.Analytical reverse diffusion step using $\alpha/\beta$ schedules, 3.Additive noise injection for non-terminal steps.

The model utilizes a hybrid approach combining Transformer-based condition processing with efficient MLP-based diffusion, suitable for tasks requiring conditional sequence generation with controllable denoising dynamics.
\section{Implementation details}
\subsection{Benchmark}
We train and test the models separately for each region. Specifically, we use the training data from a particular region to predict potential earthquakes occurring within the three months following the end of the training period and compare these predictions with the real earthuquake records from that region. We calculate the time difference between each training data entry and the previous one as an additional attribute(delta\_time). To enhance the model's predictive capability for high-magnitude earthquakes, the weight for data with magnitudes $\geq$5 is set to 10 times that of ordinary data.

Our neural network models all take four-dimensional inputs, corresponding to the following attributes in order: magnitude, latitude, longitude, and delta\_time. The output is also a four-dimensional tensor with channels matching the input. All models are configured with 196 time steps and trained for 400 epochs on an RTX 3090 GPU. The batch sizes are set to 32 for EP-CNN, EP-LSTM, and EP-Transformer, while EP-Diff uses a batch size of 64. These models are trained under  a weighted MSE loss, where the weights for longitude and latitude are increased by a factor of 3 to strengthen the model's spatial prediction capability.
\subsection{Moment Tensor Experiment}
We conducted comparative experiments on the moment tensor information using three models: EP-LSTM, EP-Transformer, and EP-Diff.  Specifically, for training and prediction, the input and output tensors of these three models were converted to 16 dimensions. These 16 dimensions represent the following information: 'magnitude', 'latitude', 'longitude', 'delta\_time', 'np1\_strike', 'np1\_dip', 'np1\_rake', 'np2\_strike', 'np2\_dip', 'np2\_rake', 't\_plunge', 't\_azimuth', 'n\_plunge', 'n\_azimuth', 'p\_plunge', 'p\_azimuth'. We used earthquake data with magnitudes $\geq$ 6 from the regions "JPN-IDN" and "USA-CL" between March 1, 1996, and March 1, 2021, for training, discarding incomplete data.  The prediction period is one year into the future (up to March 2022), and the predictions are compared with the real earthquakes with magnitudes $\geq$ 6 that occurred during this period. Due to limited training data, the performance of the models was affected, and the distance window for this experiment was set to 2000 km. The other settings remained unchanged.

\end{document}